\documentclass{aa}
\usepackage{graphicx}
\usepackage{hyperref}
\hypersetup{colorlinks,allcolors=blue}
\usepackage{txfonts}
\begin{document}

   \title{Polycyclic Aromatic Hydrocarbons in Exoplanet Atmospheres}

   \subtitle{I. Thermochemical Equilibrium Models}

   \author{Dwaipayan Dubey
          \inst{1,2},   
          Fabian Grübel\inst{1}, Rosa Arenales-Lope\inst{1,2}, Karan Molaverdikhani\inst{1,2}, Barbara Ercolano\inst{1,2}, Christian Rab\inst{1,3}, \and Oliver Trapp\inst{4}
          }

   \institute{Universitäts-Sternwarte, Fakultät für Physik, Ludwig-Maximilians-Universität München, Scheinerstr. 1, D-81679 München, Germany\\
              \email{ddubey@usm.lmu.de}
              \and Exzellenzcluster `Origins’, Boltzmannstr. 2, D-85748 Garching, Germany
         \and 
            Max-Planck-Institut für Extraterrestrische Physik, Giessenbachstr. 1, 85748 Garching, Germany
         \and
             Fakultät für Chemie und Pharmazie, Ludwig-Maximilians-Universität München, Butenandtstr. 5–13, 81377 München, Germany
             }

   \date{Received; accepted }

% \abstract{}{}{}{}{} 
% 5 {} token are mandatory
 
  \abstract
   {Polycyclic Aromatic Hydrocarbons, largely known as PAHs, are widespread in the universe and have been identified in a vast array of astronomical observations from the interstellar medium to protoplanetary discs. They are likely to be associated with the chemical history of the universe and the emergence of life on Earth. However, their abundance on exoplanets remains unknown.}
   {We aim to investigate the feasibility of PAH formation in the thermalized atmospheres of irradiated and non-irradiated hot Jupiters around Sun-like stars.}
   {To this aim, we introduced PAHs in the 1-D self-consistent forward modeling code petitCODE. We simulated a large number of planet atmospheres with different parameters (e.g. carbon to oxygen ratio, metallicity, and effective planetary temperature) to study PAH formation. By coupling the thermochemical equilibrium solution from petitCODE with the 1-D radiative transfer code, petitRADTRANS, we calculated the synthetic transmission and emission spectra for irradiated and non-irradiated planets, respectively, and explored the role of PAHs on planet spectra.}
   {Our models show strong correlations between PAH abundance and the aforementioned parameters. In thermochemical equilibrium scenarios, an optimal temperature, elevated carbon to oxygen ratio, and increased metallicity values are conducive to the formation of PAHs, with the carbon to oxygen ratio having the largest effect.}
   {}

   \keywords{planetary atmosphere --
                chemistry --
                PAHs
               }
\authorrunning{Dwaipayan Dubey et al.}
   \maketitle
%
%-------------------------------------------------------------------
\section{Introduction}

   Polycyclic Aromatic Hydrocarbons (PAHs) are an essential and pervasive constituent of carbonaceous materials in space. It has been two decades since the PAH hypothesis was proposed to explain the association of unidentified IR bands (UIRs) with interstellar environments (ISM). Though 5-20\% of ISM carbon is locked up in PAHs (Joblin \& Tielens \citeyear{joblin2011pahs}), understanding the formation and evolution of such complex organic structures is still challenging for many astrophysical settings. PAHs still remain one of the most researched chemical species in different branches of science (e.g. astrophysics, chemistry, atmospheric science, etc.).   

PAHs exist as an assortment of radicals, ions, and neutral entities in the ISM (Allamandola et al. \citeyear{allamandola1999modeling}). They are expected to have a significant influence on the chemical and hydrodynamical evolution of protoplanetary discs as well as on the dynamics of atmospheric structures of newborn exoplanets (Ercolano et al. \citeyear{10.1093/mnras/stac505}; Gorti et al. \citeyear{Gorti_2009}). They are essential for understanding the ionization equilibrium of the medium and, consequently, the evolution of gaseous atmospheres (Thi et al. \citeyear{2019A&A...632A..44T}). Moreover, from an astrobiological standpoint, PAHs are considered to be associated with pre-biotic chemistry and abiogenesis, constituting an essential step in the production of amino acids and nucleotides (Ehrenfreund et al. \citeyear{ehrenfreund2006experimentally}, \citeyear{EHRENFREUND2007383}; Rapacioli et al. \citeyear{rapacioli2006formation}; Ehrenfreund \& Charnley \citeyear{ehrenfreund2000organic}; Wakelam \& Herbst \citeyear{wakelam2008polycyclic}; Galliano \& Dwek \citeyear{galliano2008stellar}; Kim et al. \citeyear{kim20123}; Sandstrom et al. \citeyear{Sandstrom_2012}; Puzzarini et al. \citeyear{puzzarini2017spectroscopic}). 

Furthermore, PAHs show interesting photochemical properties and can act as photosensitizers and photoinitiators in numerous chemical reactions, such as polymerizations. This is highly interesting in the context of prebiotic chemistry, where PAHs can activate prebiotically formed photoredox organocatalysts such as imidazolidine-4-thiones (Closs et al. \citeyear{closs2020prebiotically}) and enable chemical transformations such as the alpha-alkylation of aldehydes, thus providing precursors of canonical amino acids. Such processes bridge the gap between the formation of small molecular building blocks, e.g. obtained by meteorite catalysis, and the formation of more complex precursors for the assembly of amino acids and nucleosides. Such a system represents the first photochemical process, utilizing the sun light as an energy source for chemical energy. In addition, the first evolutionary steps in these photoredox organocatalysts can be initiated at the molecular level (Closs et al. \citeyear{closs2022dynamic}).

Unraveling the formation and subsequent growth history of complex organic molecules has become the central focus of the PAH community in recent times. The fundamental physicochemical mechanisms of the formation process of large PAHs remain obscure, especially with regard to the important step of chemical reactions between tiny organic components and interstellar dust (Hanine et al. \citeyear{Hanine_2020}). Molecular dynamics, quantum chemistry, and laboratory experiments have suggested different chemical pathways leading to their synthesis in thermalized conditions from different precursor molecules (Hanine et al. \citeyear{Hanine_2020}; Hirai \citeyear{hirai2021molecular}; He et al. \citeyear{he2023optical}; Campisi et al. \citeyear{campisi2021interstellar}). Moreover, photochemistry and vertical mixing might also play an important role in determining their abundance in planetary atmospheres. However, their significance has not been robustly explored yet in any of the atmospheric models.

Ercolano et al. (\citeyear{10.1093/mnras/stac505}) have shown the 3.3 $\mu$m feature (aromatic C-H stretching vibrational mode) to be a promising option for the detection of PAHs on planet atmospheres and also, on protoplanetary discs. While PAHs have been detected substantially on Earth and Saturn's moon Titan (Dinelli et al. \citeyear{https://doi.org/10.1002/grl.50332}; López-Puertas et al. \citeyear{López-Puertas_2013}), the hunt for PAHs on exoplanets still remains elusive due to the limitation of ground and space-based telescopes: (a) the 3.3 $\mu$m atmospheric window on Earth is contaminated by other molecular fingerprints (Seok \& Li \citeyear{Seok_2017}) and (b) previous space-based systems lacked the required wavelength coverage. The James Webb Space Telescope (JWST), specifically the NIRSpec PRISM mode (0.6-5.3 $\mu$m), presents a solution to these shortcomings. The upcoming space missions, Ariel (Tinetti et al. \citeyear{tinetti2018chemical}, \citeyear{ec7e771e53fa417abdd3fd7c5cea7b88}), and Twinkle (Edwards et al. \citeyear{edwards2019exoplanet}), are also suitable to observe at 3.3 $\mu$m with high sensitivity (Ercolano et al. \citeyear{10.1093/mnras/stac505}).

 In this paper, we aim to present a preliminary study of PAH formation under thermalized circumstances. Our main focus is to understand the influence of different parameters on PAH production. We self-consistently simulated a large grid of planetary atmospheres for some irradiated hot Jupiters and directly imaged planets and obtained synthetic spectra for different conditions. Our procedures and assumptions are outlined in Section \ref{sec:methods}. In the next section (i.e. Section \ref{sec:results}), we present the results from our atmospheric models. Section \ref{sec:conclusion} concludes with a concise summary of our findings.

%%%%%%%%%%%%%%%%%%%%%%%%%%%%%%%%%%%%%%%%%%%%%%%%%%%%%%%%%%%%%%%%%%%
%%%%%%%%%%%%%%%%%%%%%%%%%%%%%%%%%%%%%%%%%%%%%%%%%%%%%%%%%%%%%%%%%%%
%%%%%%%%%%%%%%%%%%%%%%%%%%%%%%%%%%%%%%%%%%%%%%%%%%%%%%%%%%%%%%%%%%%
\section{Methods}
\label{sec:methods}
We have produced a grid of self-consistent planetary atmospheres for irradiated and non-irradiated hot Jupiters with  petitCODE (Molli\`ere et al. \citeyear{Mollière_2015}, \citeyear{molliere2017modeling}) to study the effect of different parameters on PAH formation under thermalized conditions. In the following sections,  petitCODE is described along with the parameter space and models.

\subsection{petitCODE}
\label{sec:petitcode}

 petitCODE is a 1-D self-consistent model (Molli\`ere et al. \citeyear{Mollière_2015}, \citeyear{molliere2017modeling}) that solves the atmospheric structures for some physical input parameters and under the assumption of radiative-convective equilibrium (both absorption/emission and scattering are considered). After certain iterations of the atmospheric structure, the final solution of the model is obtained when chemical abundances, their opacities, and radiation fields are in equilibrium with the temperature-pressure structure (TP structure) of the atmosphere. Equilibrium chemical abundances are computed using  petitCODE's self-written Gibbs free energy minimizer.  petitCODE uses the Feautrier method to obtain the radiative transfer solution and applies the Accelerated Lambda Iteration (Olson et al. \citeyear{OLSON1986431}) and Ng acceleration (Ng \citeyear{doi:10.1063/1.1682399}) method to speed up convergence.

The following physical parameters are used as inputs for the code:  effective stellar temperature, stellar radius, star-planet distance or effective planetary temperature, planet mass and radius (or surface gravity), atomic elemental abundances, and preferred insolation treatment (day-side averaged or globally averaged or angle of incidence for irradiation).

There are also options for cloud treatment inside planet atmosphere: providing the lognormal particle distribution width ($\sigma\mathrm{_g}$), the
settling factor ($f\mathrm{_{sed}}$), and vertical mixing ($K_\mathrm{{zz}}$) (following the Ackerman \& Marley \citeyear{Ackerman_2001} prescription); or by
introducing the size of cloud particles and setting the maximum
cloud mass fraction (see Mollière et al. \citeyear{molliere2017modeling}). In our present investigation, we focus on irradiated and non-irradiated hot Jupiters under cloud-free conditions, thus excluding considerations of condensation processes in our models. Further dedicated  studies will be needed to explore cloudy planet atmospheres.

Atomic species, their mass fractions, and their product molecules (gaseous, condensate, and liquid species) are needed for the calculation (see Molaverdikhani et al. \citeyear{Molaverdikhani_2019} for details). Additionally, the inclusion or non-inclusion of collision induced absorption (CIA) species in the models needs to be specified. 

%%%%%%%%%%%%%%%%%%%%%%%%%%%%%%%%%%%%%%%%%%%%%%%%%%%%%%%%%%%
%%%%%%%%%%%%%%%%%%%%%%%%%%%%%%%%%%%%%%%%%%%%%%%%%%%%%%%%%%%
\subsection{Grid Properties}
\label{sec:Grid}

We explore a grid of models with varying carbon to oxygen ratio, metallicity, planetary effective temperature (see section \ref{sec:0D_Model}), and opacity species. We limit our study to the case of a G5 star ($T_*$ = 5660 K, $R_*$ = 1$R_{\odot}$, V = 9, and K = 7.5) and log($g$) = 3 ($R_\mathrm{p}$ = 1$R_\mathrm{J}$, $M_\mathrm{p}$ = 1$M_\mathrm{J}$ where $R_\mathrm{J}$ and $M_\mathrm{J}$ are the radius and mass of Jupiter respectively).

\subsubsection{Carbon to Oxygen Ratio (C/O)}
\label{sec:ctoo}

The C/O ratio can have a significant influence on the TP structure of the atmosphere and the distribution pattern of molecular abundances in atmospheric layers. Photochemical hazes are anticipated to be important for exoplanet atmospheres, especially those with higher C/O ratios or higher metallicities (He et al. \citeyear{he2023optical}; Hörst et al. \citeyear{2018NatAs...2..303H}). For this reason, we focused on super-solar C/O ratios, starting from solar value: C/O = [0.55, 0.80, 1.00, 1.25, 1.50].

We varied the C/O ratio according to the recipe chosen by Madhusudhan (\citeyear{Madhusudhan_2012}), Molli\`ere et al. (\citeyear{Mollière_2015}), Woitke et al. (\citeyear{2018A&A...614A...1W}), and Molaverdikhani et al. (\citeyear{Molaverdikhani_2019}): by changing the oxygen elemental abundance while keeping the carbon elemental abundance constant. This captures the accumulation of variable water content on a young accreting planet through gases and planetesimals.

%%%%%%%%%%%%%%%%%%%%%%%%%%%%%%%%%%%%%%%%%%%%%%%%%%%%%%%%%%%%%%%%%
\subsubsection{Metallicity [Fe/H]}
\label{sec:metallicity}

As mentioned in section \ref{sec:ctoo}, photochemical hazes may have a crucial impact on atmospheres in the presence of enhanced metallicity values. Gas and ice giants in our solar system show diversified bulk composition, ranging from 3x to 100x the solar metallicity value (Guillot et al. \citeyear{guillot2022giant}). In this work, we choose to explore an extended range of metallicities in our thermochemical models spanning from solar to 10000x solar. While a metallicity of 10000x solar is unrealistic for any gas or ice giant, it is useful to provide a limiting case for PAH formation.

Our metallicity grid spans the following values: [Fe/H] = [0, 1, 2, 3, 4]. Here, [Fe/H] stands for metallicity in the log scale. So, [Fe/H] = 1 signifies a 10 times higher metal-rich atmosphere than our Sun. All elements except H and He are considered as ``metal'' here.

%%%%%%%%%%%%%%%%%%%%%%%%%%%%%%%%%%%%%%%%%%%%%%%%%%%%%%%%%%%%%%%%%%%%%
\subsubsection{Reactant and Opacity Species}
\label{sec:opacity}

In this work, we employ  petitCODE's default chemical network, which includes atomic species and their products. The following gas opacity species are provided for the models: $\mathrm{CH_4}$, $\mathrm{H_2O}$, $\mathrm{CO_2}$, $\mathrm{HCN}$, $\mathrm{CO}$, $\mathrm{H_2S}$, $\mathrm{NH_3}$, $\mathrm{C_2H_2}$, $\mathrm{Na}$, $\mathrm{K}$, $\mathrm{TiO}$, and $\mathrm{VO}$. CIA species ($\mathrm{H_2}$-$\mathrm{H_2}$ and $\mathrm{H_2}$-$\mathrm{He}$) are also taken into account.

In the modeling of directly imaged planets, some condensate species ($\mathrm{Mg_2SiO_4}$(c), $\mathrm{MgAl_2O_4}$(c), and $\mathrm{Fe_2SiO_4}$(c)) had to be removed from the network due to poor convergence of the models. This is a known problem, lying with the abundance matrix in Gibbs's free energy minimizer (see Appendix A of Molaverdikhani et al. \citeyear{Molaverdikhani_2019} for more details).

%%%%%%%%%%%%%%%%%%%%%%%%%%%%%%%%%%%%%%%%%%%%%%%%%%%%%%%%%%%%%%%%%%%%%%%%%%
%%%%%%%%%%%%%%%%%%%%%%%%%%%%%%%%%%%%%%%%%%%%%%%%%%%%%%%%%%%%%%%%%%%%%%%%%%
\subsection{0-D Models for organic chemical species}
\label{sec:0D_Model} % used for referring to this section from elsewhere

As predicted by molecular dynamics simulations (Hanine et al. \citeyear{Hanine_2020}; Hirai \citeyear{hirai2021molecular}), PAH formation is favored at a critical temperature region. Lower temperatures are insufficient for PAHs to form thermally, and higher temperatures lead to the thermal cleavage of PAHs. To check the critical temperature inside the planet's atmosphere, we ran a set of 0-D models for 180 organic species (alkanes, alkenes, alkynes, and aromatic species, etc.) with the equilibrium chemistry solver of  petitCODE for a C/O ratio of 1.25. NASA9 polynomial coefficients (for Gibbs free energy calculation) for all the molecules have been taken from the Burcat (\citeyear{Burcat}) database.

We considered a wide range of temperatures (800K, 1000K, 1100K, 1200K, 1300K, 1400K, 1500K, 1600K, 1750K, and 2000K) and fixed the pressure at 10$^{-2}$ bar for the calculation. The temperature grid was designed to encompass the full range of observed planetary effective temperatures ($T_{\mathrm{eff}}$), excluding extremely irradiated hot Jupiter, which might have intrinsically different atmospheric chemistry (Molaverdikhani et al. \citeyear{Molaverdikhani_2019}). The pressure layer was also chosen in terms of planetary photosphere context. The photosphere is the specific region inside a planet's atmosphere (between 1-10$^\mathrm{-3}$ bar) that we probe for the transmission and emission spectra of a planet.

\begin{figure*}
    \centering  
        \begin{minipage}[b]{\columnwidth}
            \includegraphics[width=\columnwidth]{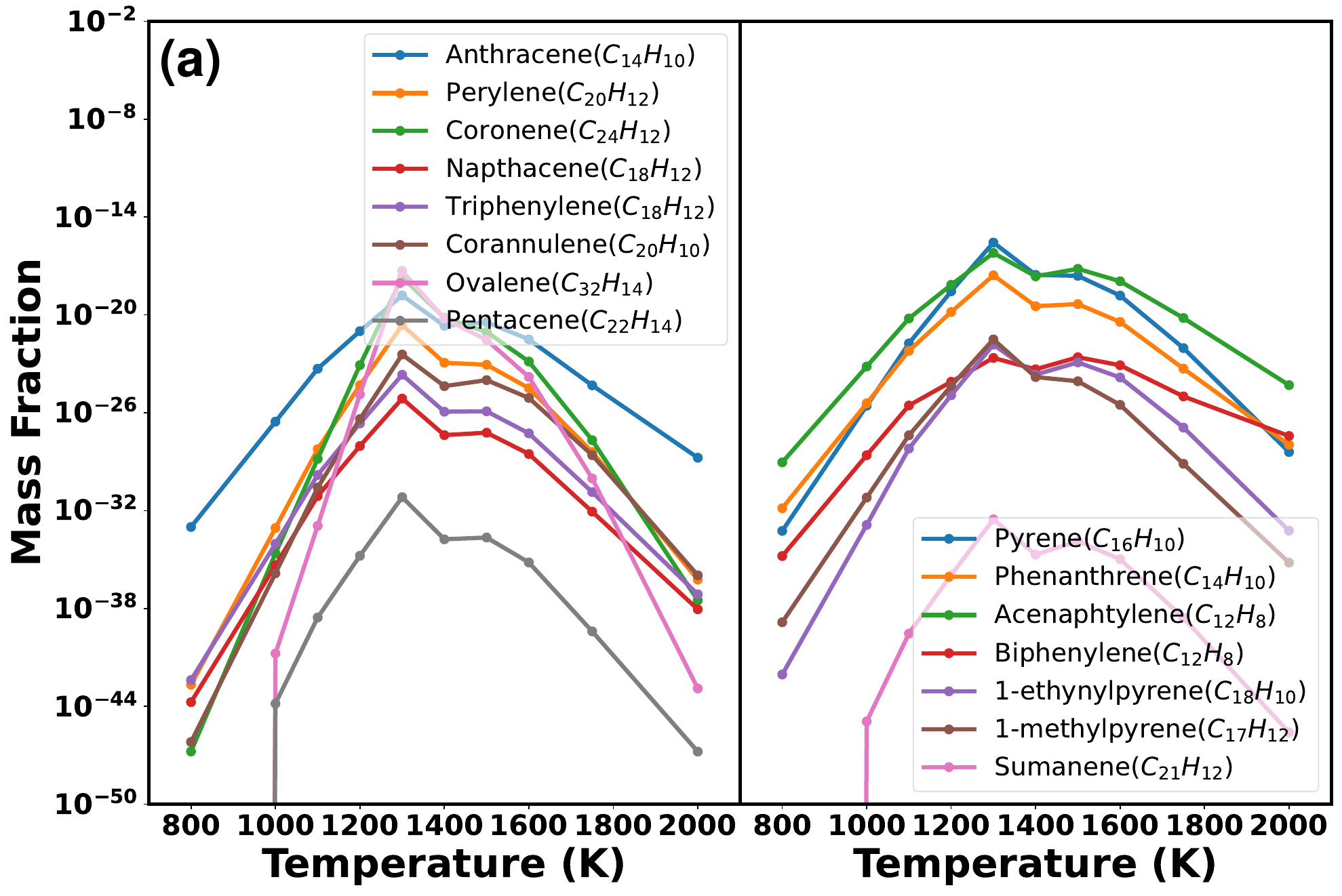}
        \end{minipage}
        %\columnbreak
         \begin{minipage}[b]{\columnwidth}
            \includegraphics[width=\columnwidth]{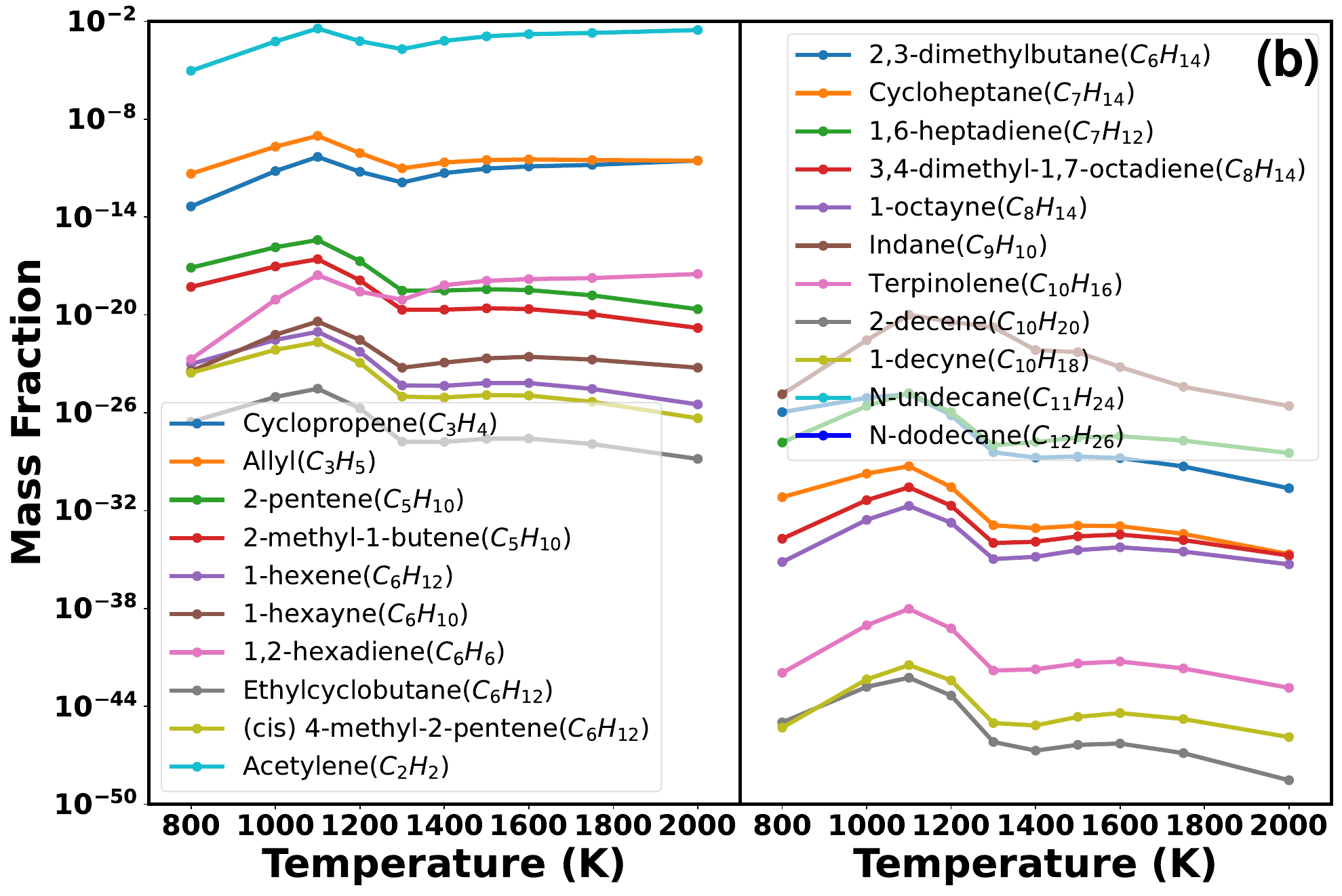}
        \end{minipage}
        %\columnbreak
         \begin{minipage}[b]{\columnwidth}
            \includegraphics[width=\columnwidth]{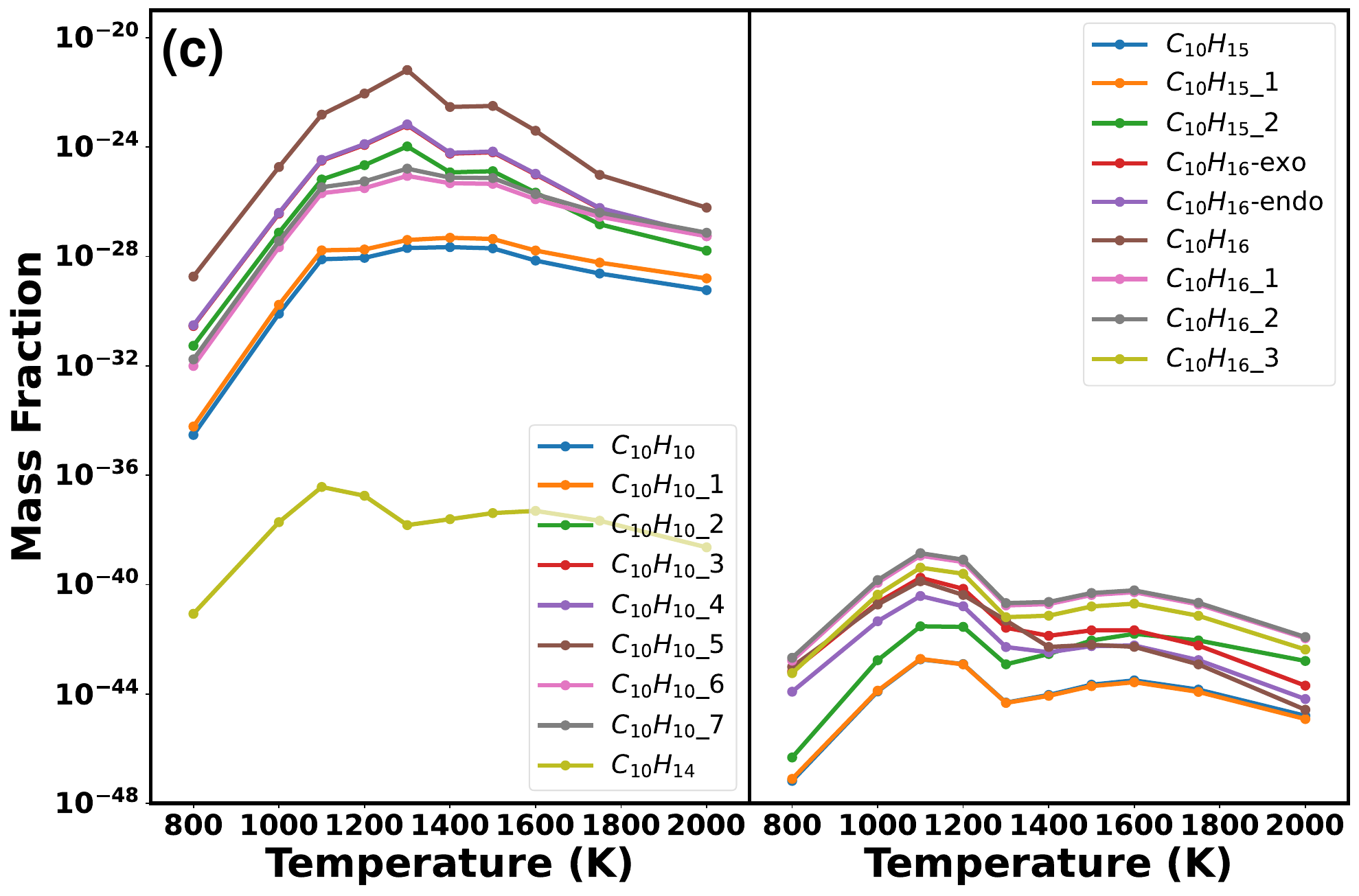}
        \end{minipage}
        %\columnbreak
         \begin{minipage}[b]{\columnwidth}
            \includegraphics[width=\columnwidth]{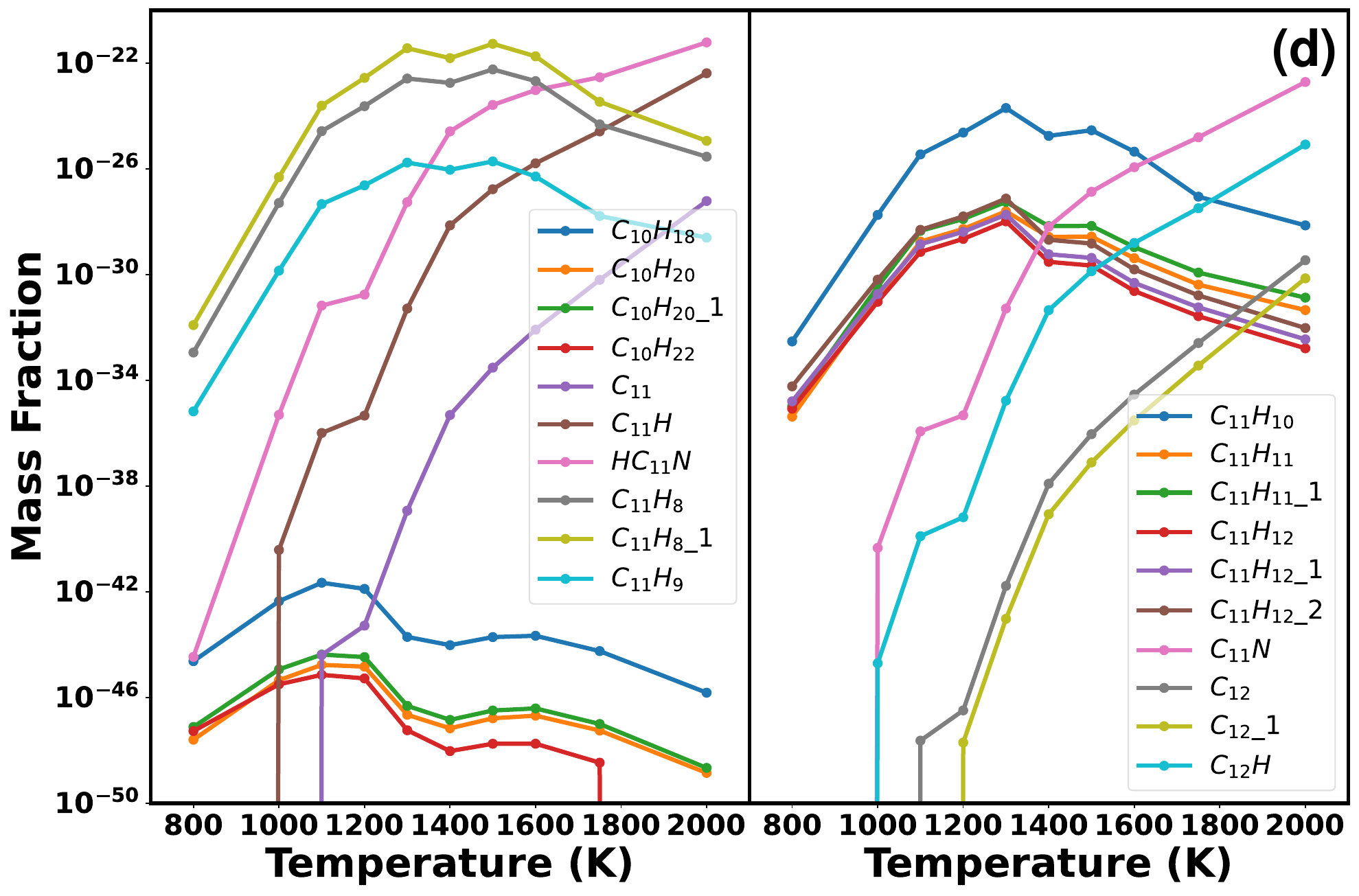}
        \end{minipage}
         \begin{minipage}[b]{\columnwidth}
            \includegraphics[width=\columnwidth]{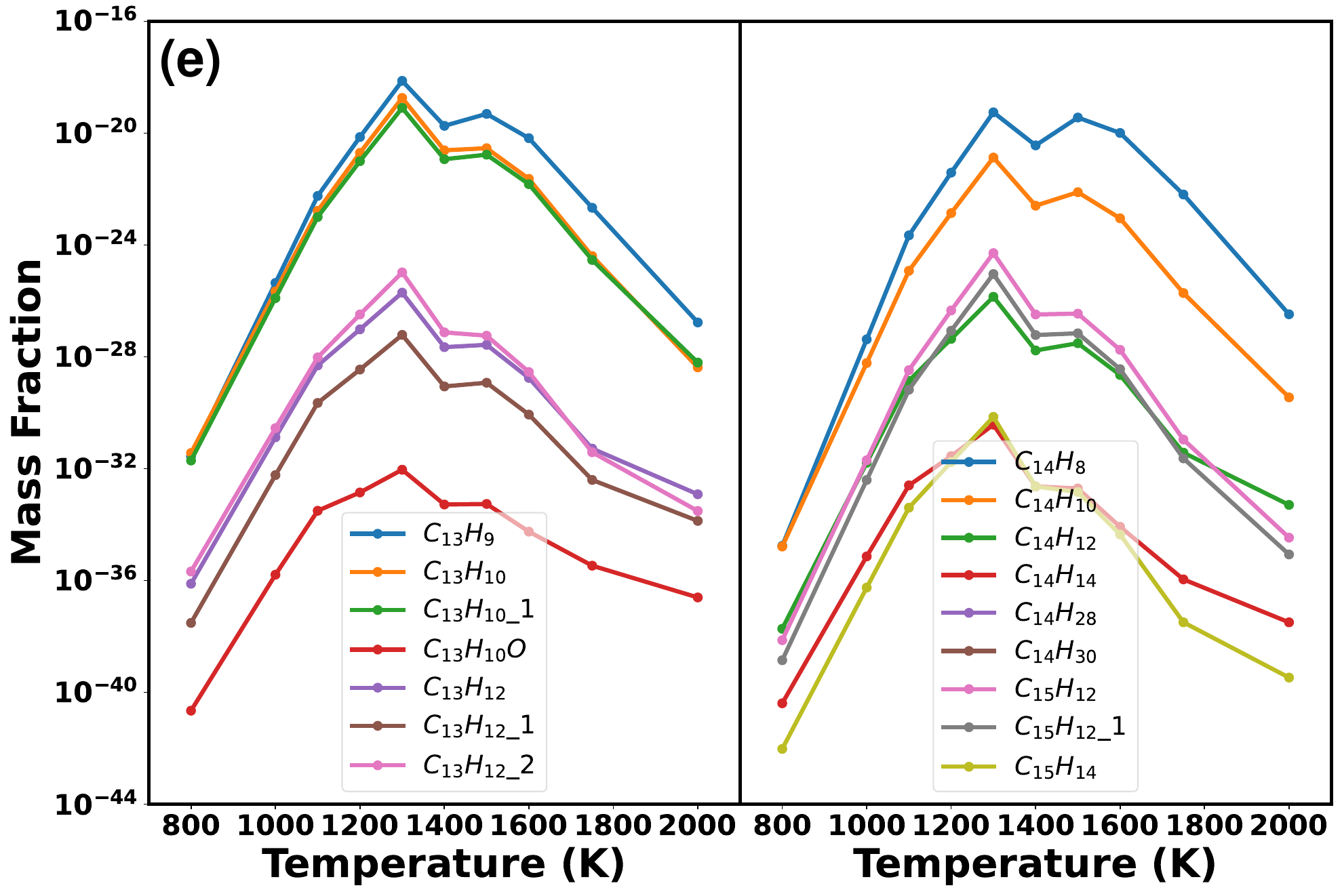}
        \end{minipage}
\caption{0-D modeling of 90 organic species (PAHs and other organic molecules) for P = $\mathrm{10^{-2}}$ bar and over a wide temperature grid:  The $n\mathrm{^{th}}$ isomer for some species is displayed with the symbol `$\mathrm{\_ n}$' (n is an integer). Subfigures (a) and (b) showcase the PAHs and saturated/less unsaturated species, respectively. Entities with more unsaturation are separately shown in (e). (c) and (d) display a combination of different categories together.}
\label{fig:0D_model}
\end{figure*}

%%%%%%%%%%%%%%%%%%%%%%%%%%%%%%%%%%%%%%%%%%%%%%%%%%
\subsection{Modelling planets for PAH formation}
\subsubsection{1-D Self-consistent Models for transiting planets}
\label{sec:self-consistent} % used for referring to this section from elsewhere

Based on the result from the 0-D models (see Figure \ref{fig:0D_model}), we calculated self-consistent forward models of planetary atmospheres with $T_{\mathrm{eff}}$ = 1300K, planet internal temperature ($T_{\mathrm{int}}$) = 200K, but with different metallicity and C/O ratios. 
 We considered $T_{\mathrm{int}}$ $\ll$ $T_{\mathrm{eff}}$ to satisfy the following condition for irradiated planets: $T_{\mathrm{eqm}}$ $\approx$ $T_{\mathrm{eff}}$. This can easily be understood from a simple relation: $T_{\mathrm{eff}}^4$ = $T_{\mathrm{eqm}}^4$ + $T_{\mathrm{int}}^4$, where $T_{\mathrm{eqm}}$ is the planet's equilibrium temperature. 
 
For the sake of simplicity, we limited our study to a few selected PAH species: Anthracene ($C_{14}H_{10}$), Perylene ($C_{20}H_{12}$), Coronene ($C_{24}H_{12}$), and Ovalene ($C_{32}H_{14}$). These were selected because of their larger abundance predicted by thermochemical equilibrium models (see Section \ref{result:0D_model}). 

%%%%%%%%%%%%%%%%%%%%%%%%%%%%%%%%%%%%%%%%%%%%%%%%%%
\subsubsection{Extraplanetary origins of PAHs}
\label{sec:ISM} % used for referring to this section from elsewhere

PAHs are abundant species in the ISM, with a typical number density of 3x10$^{\mathrm{-7}}$ relative to H (Tielens \citeyear{tielens2008interstellar}). In addition to the in-situ formation of PAHs within a planetary atmosphere, there are several potential pathways through which a planet can accumulate PAHs at levels comparable to those found in the ISM. One significant mechanism involves the continuous influx of interplanetary dust via various astronomical processes, such as comet and asteroid impacts, among others. These events contribute to the enrichment of PAHs on the planet's surface. We convert the ISM value for the number density of PAHs to a mass fraction as follows: 

\begin{align*}
    m_\mathrm{{PAH}} \approx 3\times10^{-7}\cdot n_\mathrm{H}\cdot \sum_i M_\mathrm{{PAH}}^\mathrm{i}
\end{align*}
\begin{align*}
    M_\mathrm{{total}} \approx n_\mathrm{H}\cdot 1u + n_\mathrm{He}\cdot 4u + n_\mathrm{H_2O}\cdot 18u + + n_\mathrm{CH_4}\cdot 16u + n_\mathrm{CO}\cdot 28u \\ 
    + n_\mathrm{HCN}\cdot 27u + n_\mathrm{CO_2}\cdot 44u + \dots
\end{align*}
\begin{align}
    \mathrm{Mass \hspace{1mm} Fraction \hspace{1mm} of \hspace{1mm} PAH} = \frac{m_\mathrm{PAH}}{M_\mathrm{total}}
\end{align}

where, $m_\mathrm{PAH}$ = net mass of all PAHs, $M_\mathrm{PAH}^\mathrm{i}$ = molecular mass of $i^\mathrm{th}$ PAH species, $M_\mathrm{total}$ = total mass of molecules present in the atmosphere, and $n_\mathrm{k}$ = number density of $k^\mathrm{th}$ species in the atmosphere.

We kept the same planet parameters as listed in Section \ref{sec:self-consistent}. However, we fixed the C/O ratio to the solar value  (C/O = 0.55) and the [Fe/H] to 1, which is relevant to the question of whether PAH signatures are detectable on planets transiting Sun-like stars. Even though several attempts have been made in the past, measuring the C/O ratios of exoplanets remains a challenge due to the poor detectability of carbon-bearing species. While transiting planets and directly imaged planets exhibit a wide range of C/O ratios, the measured values are subject to enormous uncertainty (see Hoch et al. \citeyear{Hoch_2023}). Future high-resolution data from JWST might be useful to constrain the C/O ratio of planets with greater precision in the near future.

%%%%%%%%%%%%%%%%%%%%%%%%%%%%%%%%%%%%%%%%%%%%%%%%%%%%%%%%%%%%%%%%%%
\subsubsection{1-D Self-consistent model for directly imaged planets}
\label{sec:direct_imaged} % used for referring to this section from elsewhere

We further investigated if the imprints of PAHs could be visible in a carbon-rich atmosphere of a directly imaged planet. Directly imaged planets are young planets, which have generally formed far away from their host stars. Irradiation from the central star is generally negligible and they are detected in the infrared thanks to their warm interiors. We modeled the planets with star-planet distance (d) = 10au, $T_{\mathrm{int}}$ = 700K, constant C/O = 1.2 , and with varying metallicities.

We reduced the atmospheric pressure range to 10-10$^\mathrm{-3}$ bar and decreased the temperature step inside the  petitCODE to make the convergence smoother.

%%%%%%%%%%%%%%%%%%%%%%%%%%%%%%%%%%%%%%%%%%%%%%%%%%%%%%%%%%%%%%%%%%
\subsection{Calculation of transmission and emission spectra from modeled atmospheres}
\label{sec:spectra} % used for referring to this section from elsewhere

Very little is known currently about PAHs in exoplanet atmospheres, partially due to the instrumental limitations. A dedicated search for the PAH signature in transmission and emission spectra using the atmospheric retrieval method has not been attempted yet. Here, we use our atmospheric models to produce synthetic spectra to determine whether such detection is within reach of current and future instrumentation. We coupled the outputs from  petitCODE to the radiative transfer modeling and atmospheric retrieval code  petitRADTRANS (Molli\`ere et al. \citeyear{2019A&A...627A..67M}, \citeyear{2020A&A...640A.131M}) for the calculation of synthetic transmission (for transiting planets) and emission spectra (for directly imaged planets) at a resolution of $\mathrm{\lambda/\Delta\lambda}$ = 1000. We considered the line opacity species: $\mathrm{H_2O}$, $\mathrm{CO}$ (HITEMP; see Rothman et al. \citeyear{ROTHMAN20102139}), \textbf{$\mathrm{C_2H_2}$}, $\mathrm{CH_4}$, and $\mathrm{HCN}$ (ExoMol; see Chubb et al. \citeyear{2021A&A...646A..21C}); Rayleigh opacity species: $\mathrm{H_2}$, He; CIA species: $\mathrm{H_2}$-$\mathrm{H_2}$, $\mathrm{H_2}$-$\mathrm{He}$ (see Molli\`ere et al. \citeyear{2019A&A...627A..67M} for further details). 

We followed Ercolano et al. (\citeyear{10.1093/mnras/stac505}) and  calculated the PAH cross-sections following Li \& Draine (\citeyear{2001ApJ...554..778L}) and including updates from Draine \& Li (\citeyear{draine2007infrared}) for circumcoronene PAHs consisting of 54 carbon and 18 hydrogen atoms for both neutral and ionized PAHs, as described by Woitke et al. (\citeyear{2016A&A...586A.103W}). This results in a PAH mass of 667 amu and a PAH radius of 4.87Å (Weingartner \& Draine \citeyear{2001ApJS..134..263W}). These cross-sections correspond to the optical properties of ``astro-PAHs'' and align well with astronomical observations in the ISM. These opacities are regularly employed in protoplanetary disk models including the hot irradiated atmospheres of the inner regions of disks.

%%%%%%%%%%%%%%%%%%%%%%%%%%%%%%%%%%%%%%%%%%%%%%%%%%%%%%%%%%%%%%%%%%
%%%%%%%%%%%%%%%%%%%%%%%%%%%%%%%%%%%%%%%%%%%%%%%%%%%%%%%%%%%%%%%%%%%%%%%%%%%%%
%%%%%%%%%%%%%%%%%%%%%%%%%%%%%%%%%%%%%%%%%%%%%%%%%%%%%%%%%%%%%%%%%%%%%%%%%%%%%
%%%%%%%%%%%%%%%%%%%%%%%%%%%%%%%%%%%%%%%%%%%%%%%%%%%%%%%%%%%%%%%%%%%%%%%%%%%%%
\section{Results}
\label{sec:results}

\subsection{0-D Models: Optimum temperature for PAH formation and correlation with other organic species}
\label{result:0D_model}

\subsubsection{Model for a large number of molecules as carbon-bearing species}
\label{sec:large_number_species}

We estimated the mass fractions of different organic species, including saturated (alkanes) and unsaturated (alkenes, alkynes, and PAHs) hydrocarbons as a function of temperature and at a pressure value consistent with the planetary photosphere regime (see Figure \ref{fig:0D_model}). Structural isomers were also analyzed for a few species. For the sake of simplicity, we neglected charged species and oxygen or nitrogen-substituted species particularly. 

We introduce a new parameter $\beta$ ($\beta$ = $n\mathrm{_C}/n\mathrm{_H}$, where $n\mathrm{_C}$ and $n\mathrm{_H}$ refer to the number of carbon atoms and hydrogen atoms present in a molecule) that classifies the molecules into three different classes: (1) $\beta$ $\geq$ 1 (species with more unsaturation), (2) $\beta$ $\mathrm{<}$ 1  (saturated and less unsaturated species), and (3) $\beta$ $\mathrm{\gg}$ 1 (highly unsaturated hydrocarbon radicals i.e. with more carbon atoms and fewer hydrogen atoms).

The 3 classes show different correlations with temperature. Class 1 shows a peak at 1300K and dips both ways. It implies that the formation of PAHs and other highly unsaturated species is favored at 1300K in thermalized conditions. A similar trend was earlier found by  Hirai (\citeyear{hirai2021molecular}). The temperature dependency is derived from the dependency of the thermodynamical coefficients with temperature, resulting in different values of $\mathrm{\Delta G}$. Since the formation of a given molecular species requires that $\mathrm{\Delta G}$ $<$ 0 thermodynamically, for each species, there exists a specific temperature range that favors the formation of that species. This temperature range lies in the neighborhood of 1300K for highly unsaturated species.

Class 2 molecules exhibit a distinct behavior. At lower temperatures, they follow a positive correlation with temperature. But they undergo a dip at 1300K. It manifests the contribution of saturated/less unsaturated species toward the formation of Class 1 species at the optimum temperature regime. However, it is also important to note that Class 2 hydrocarbons with $n\mathrm{_C}$ $>$ 11 fail to form (see N-undecane, N-dodecane from Figure \ref{fig:0D_model}b and $C\mathrm{_{14}}H\mathrm{_{28}}$, $C\mathrm{_{14}}H\mathrm{_{30}}$ from Figure \ref{fig:0D_model}e). Carbon atoms seem to be entrapped in smaller compounds. This can be understood by looking at the relative mass fractions from Figure \ref{fig:0D_model}(b). $\beta$-factor too has a key role here. Mass fractions of Class 2 molecules decrease with decreasing the $\beta$-factor. Even though both factors have a combined influence on the abundance of Class 2 species, the former is more significant in this instance.

 Being a hydrocarbon radical entity, we neglected the role of Class 3 (Figure \ref{fig:0D_model}(d) shows 7 of them in two different subplots) in PAH formation for equilibrium models. But their presence is important for the disequilibrium models, which will be the focus of our future work.

%%%%%%%%%%%%%%%%%%%%%%%%%%%%%%%%%%%%%%%%%%%%%%%%%%%%%%%%%%%%%%%%%%%%%%%%%%%
 \subsubsection{Only PAHs as carbon bearing species}
\label{sec:only_PAHs}

It is difficult to estimate the exact magnitude of the PAH abundances in our models. The reason is that the absolute value strongly depends on how the available carbon is distributed amongst all carbon bearing species.

\begin{figure}
	\includegraphics[width=\columnwidth]{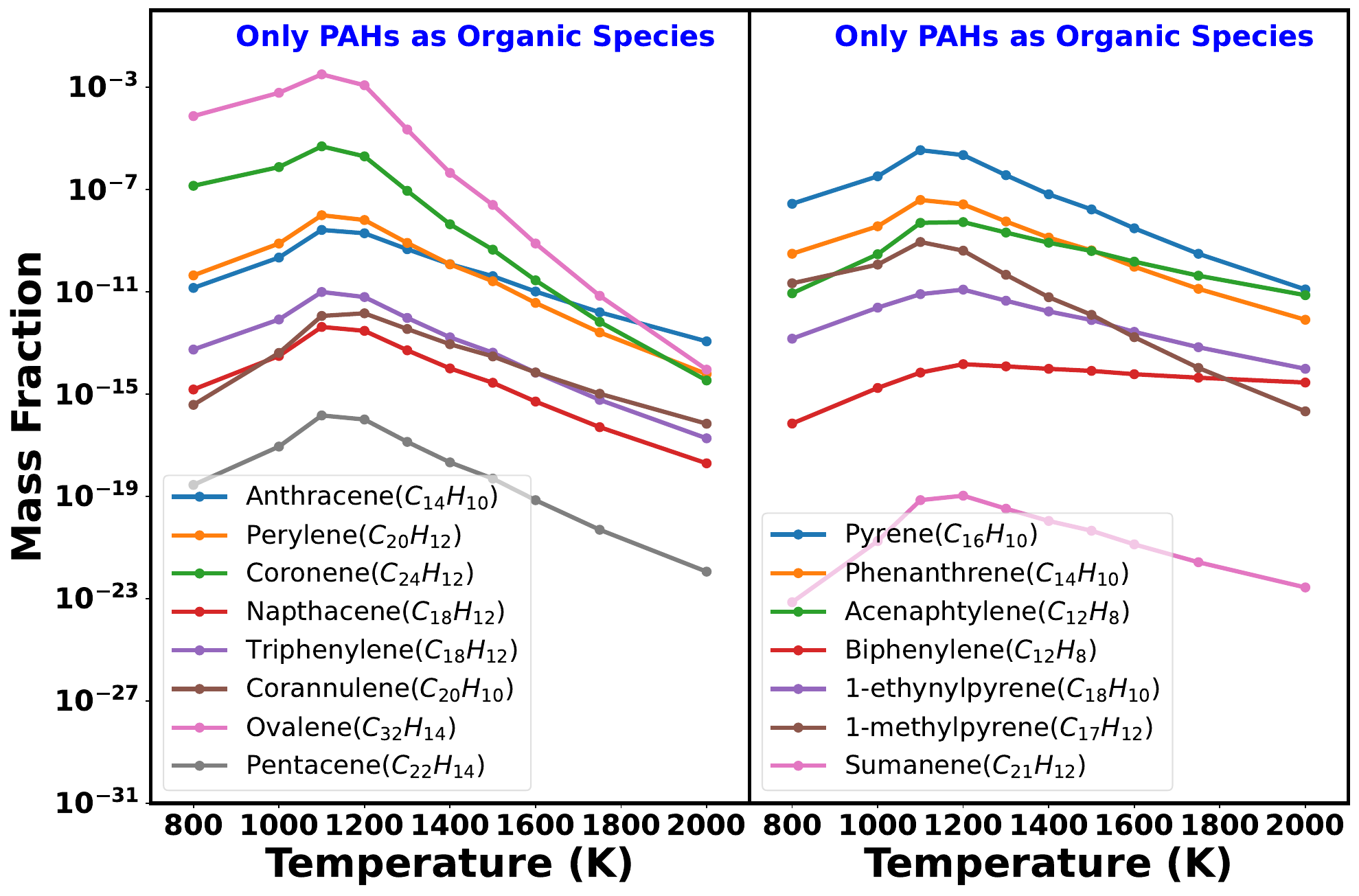}
  \caption{0-D model representing PAHs as the only carbon bearing species. Other small and large hydrocarbons have been removed from the calculation to better constrain the order of mass fractions for PAHs with temperature. }
   \label{fig:only_PAHs}
\end{figure}

An upper limit can, however, be obtained by making the assumption that all carbon available is locked into PAHs. This is shown in Figure \ref{fig:only_PAHs}, which shows PAH abundances that are several orders of magnitude higher than our previous calculation (see Figure \ref{fig:0D_model}a).

  %%%%%%%%%%%%%%%%%%%%%%%%%%%%%%%%%%%%%%%%%%%%%%%%%%%%%%%%%%%%%%%%%%%%%%%%%%%%%%%%%%%%%%%%%%%%%%%%%%%%%%%%%%%%%%%%%%%%%%%%%%%%%%%%%%%
\subsection{Constraining C/O ratio and [Fe/H] for PAH signature in planetary spectra from 1-D self-consistent models}
\label{result:spectra}

\subsubsection{Transiting planets}
\label{result:transiting_planet}

\begin{figure*}
\centering
	\includegraphics[width=2.0\columnwidth]{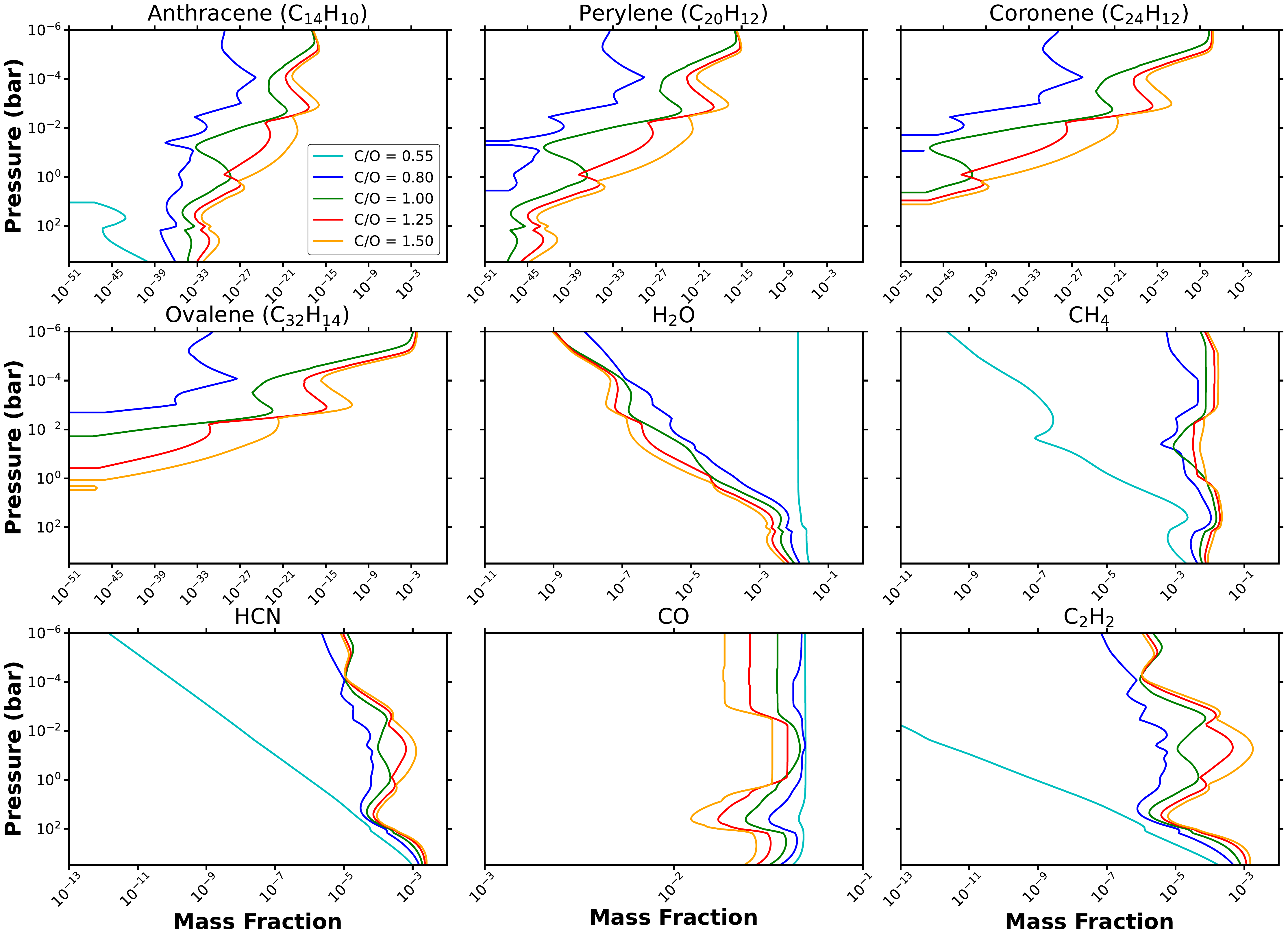}
  \caption{Mass fraction of 4 different PAHs, 
  $\mathrm{H_2O}$, $\mathrm{CH_4}$, HCN, CO, and $\mathrm{C_2H_2}$ for transiting planet ($T\mathrm{_{eff}}$ = 1300K, $T\mathrm{_{int}}$ = 200K, [Fe/H] = 1.0, and log($g$) = 3.0) for different C/O ratios.}
   \label{fig:VMR_transiting_CtoO}
\end{figure*}

\begin{figure}
	\includegraphics[width=\columnwidth]{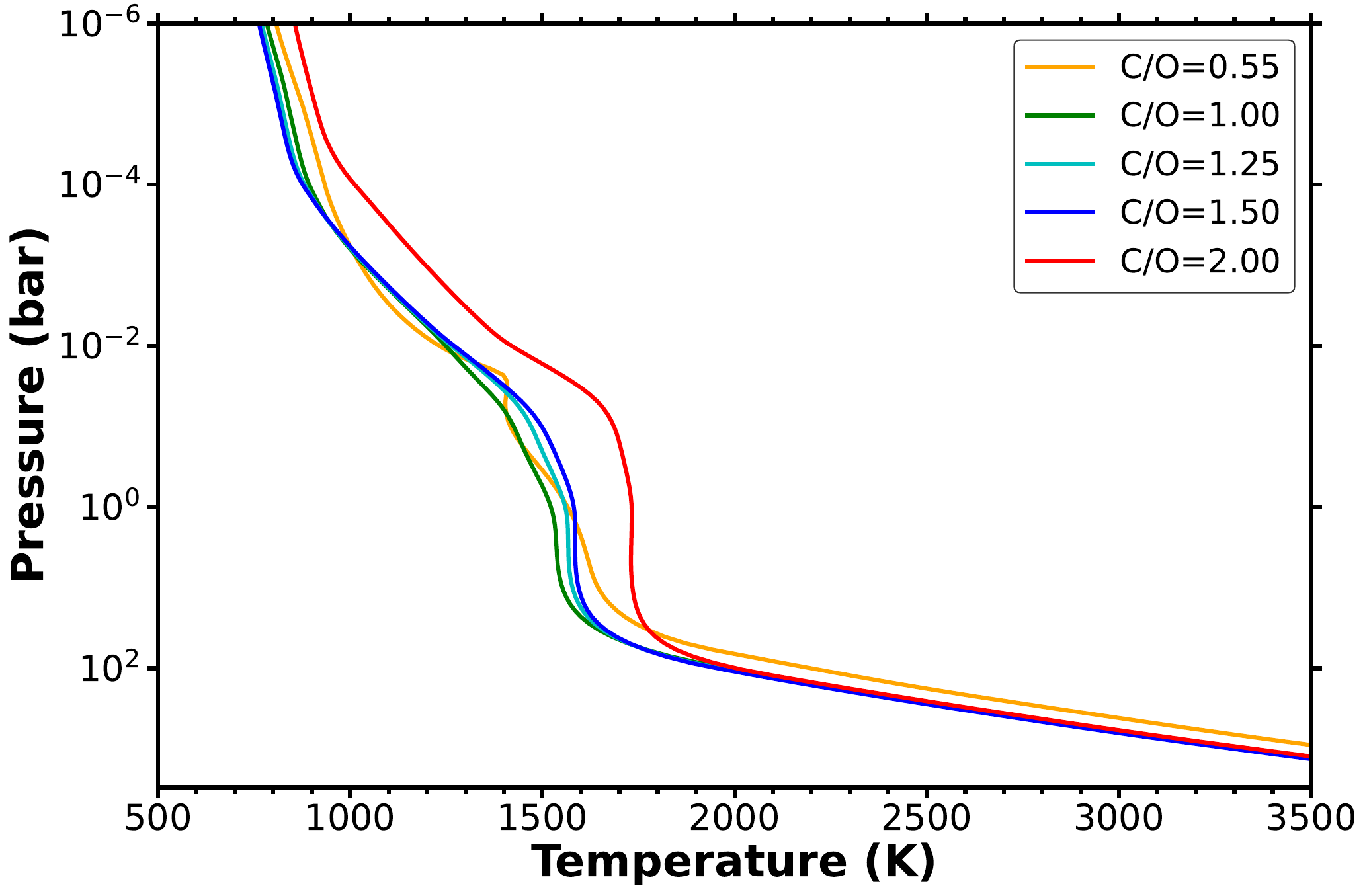}
        \caption{Pressure-Temperature profile for a modeled planet: $T\mathrm{_{eff}}$ = 1300K, $T\mathrm{_{int}}$ = 200K, [Fe/H] = 1.0, and log($g$) = 3.0.}
        \label{fig:TP_transit}
\end{figure}

\begin{figure*}
    \centering  
        \begin{minipage}[b]{\columnwidth}
            \includegraphics[width=\columnwidth]{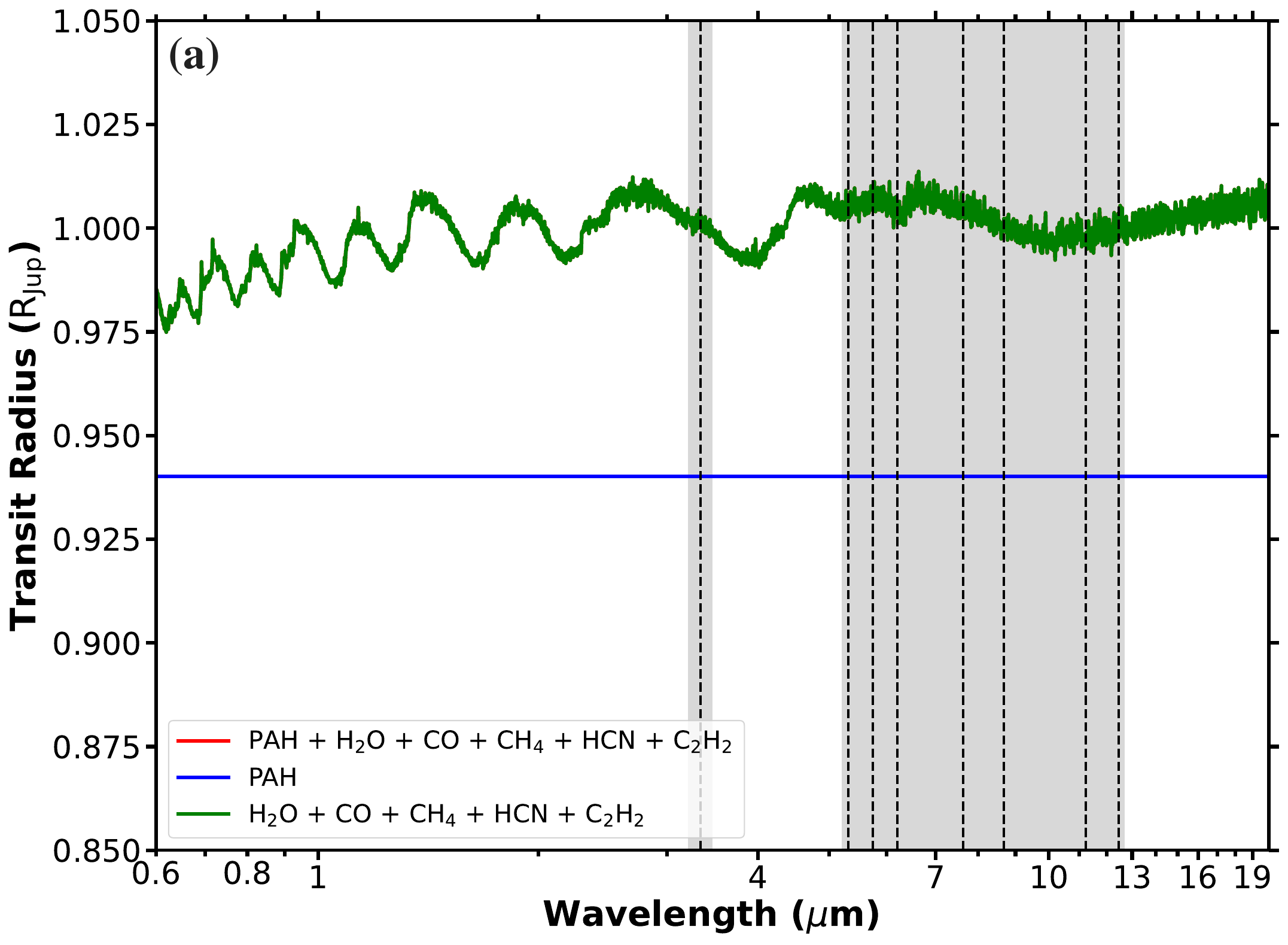}
        \end{minipage}
        %\columnbreak
         \begin{minipage}[b]{\columnwidth}
            \includegraphics[width=\columnwidth]{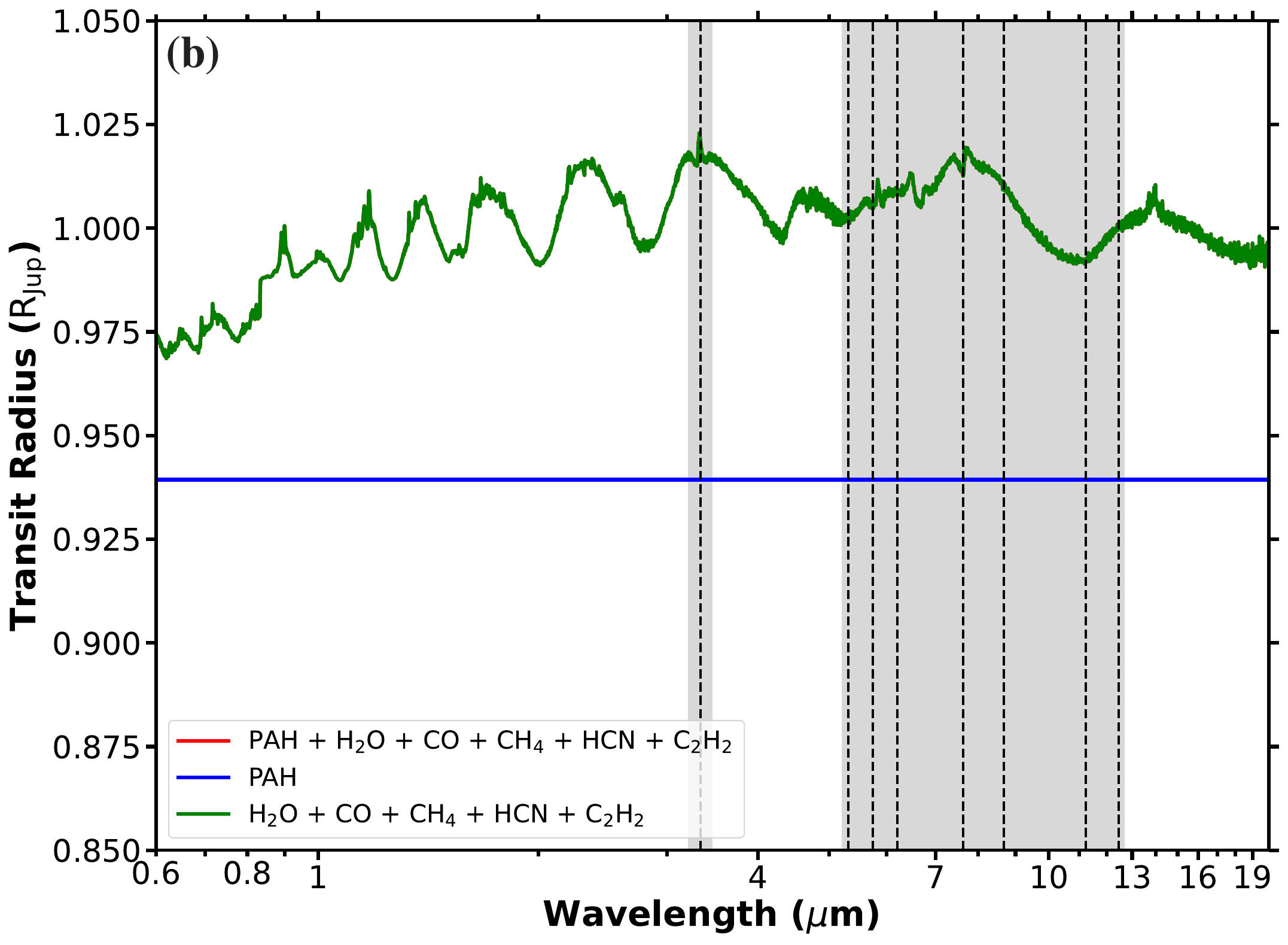}
        \end{minipage}
        %\columnbreak
         \begin{minipage}[b]{\columnwidth}
            \includegraphics[width=\columnwidth]{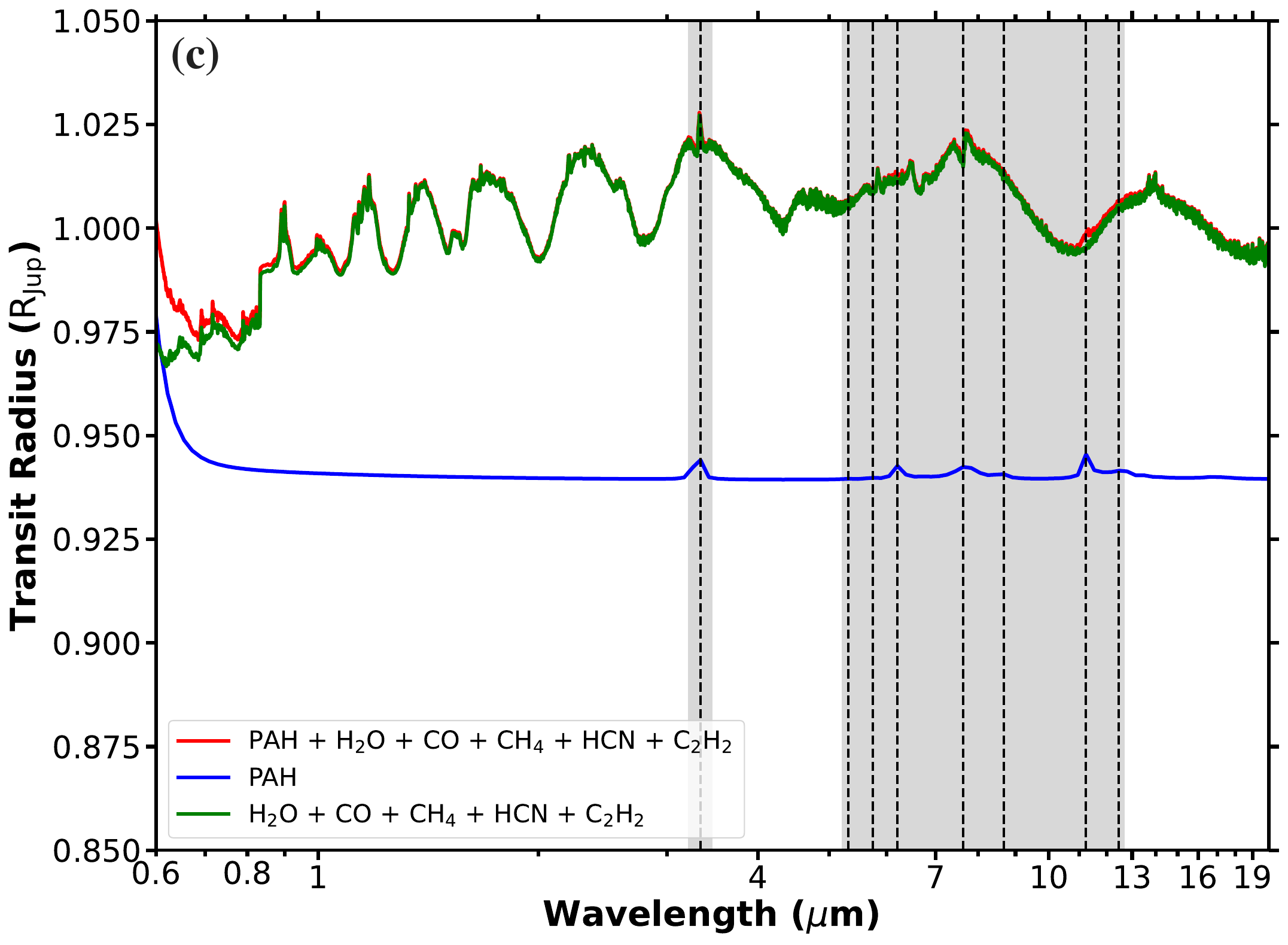}
        \end{minipage}
        %\columnbreak
         \begin{minipage}[b]{\columnwidth}
            \includegraphics[width=\columnwidth]{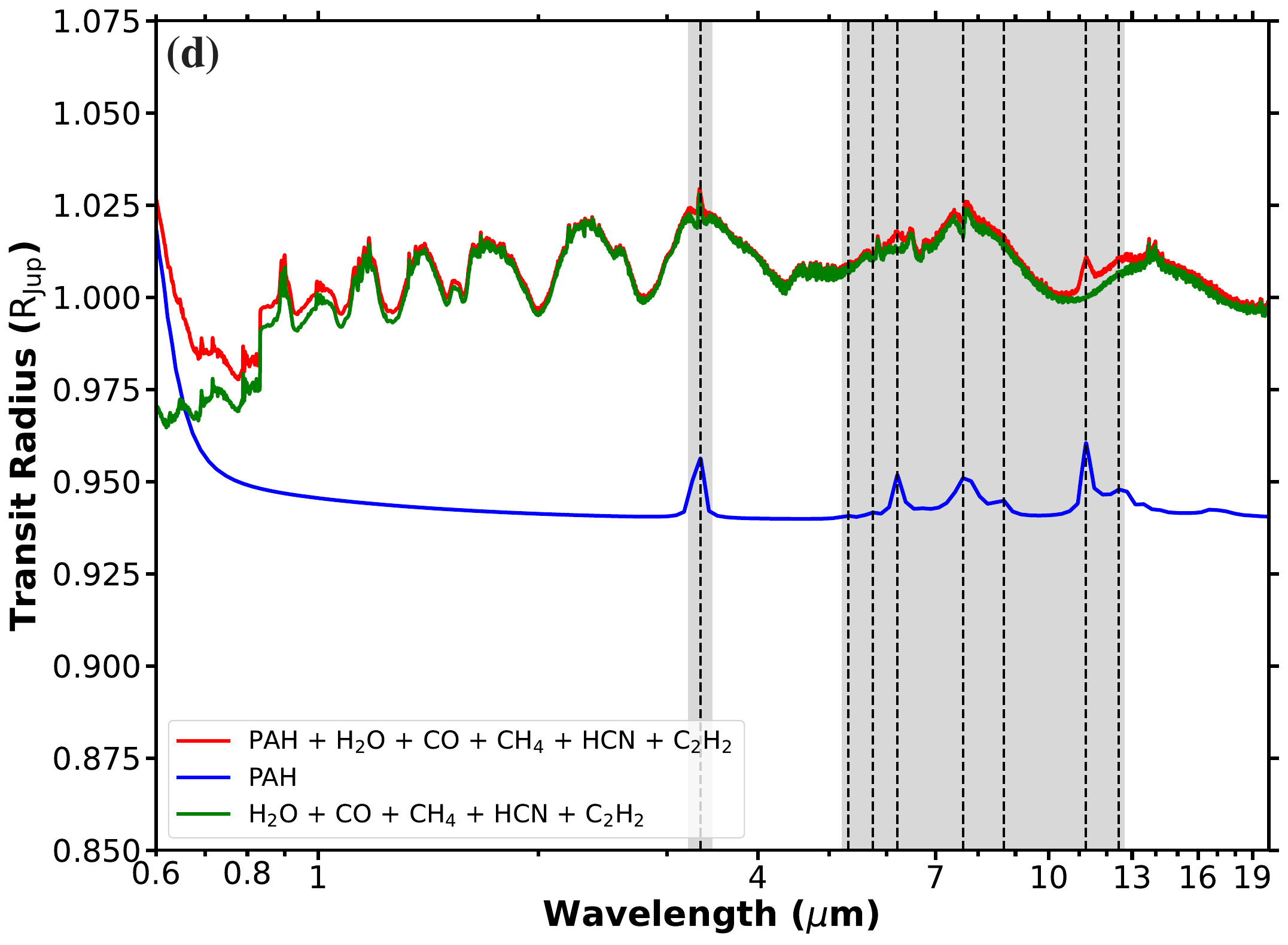}
        \end{minipage}
         \begin{minipage}[b]{\columnwidth}
            \includegraphics[width=\columnwidth]{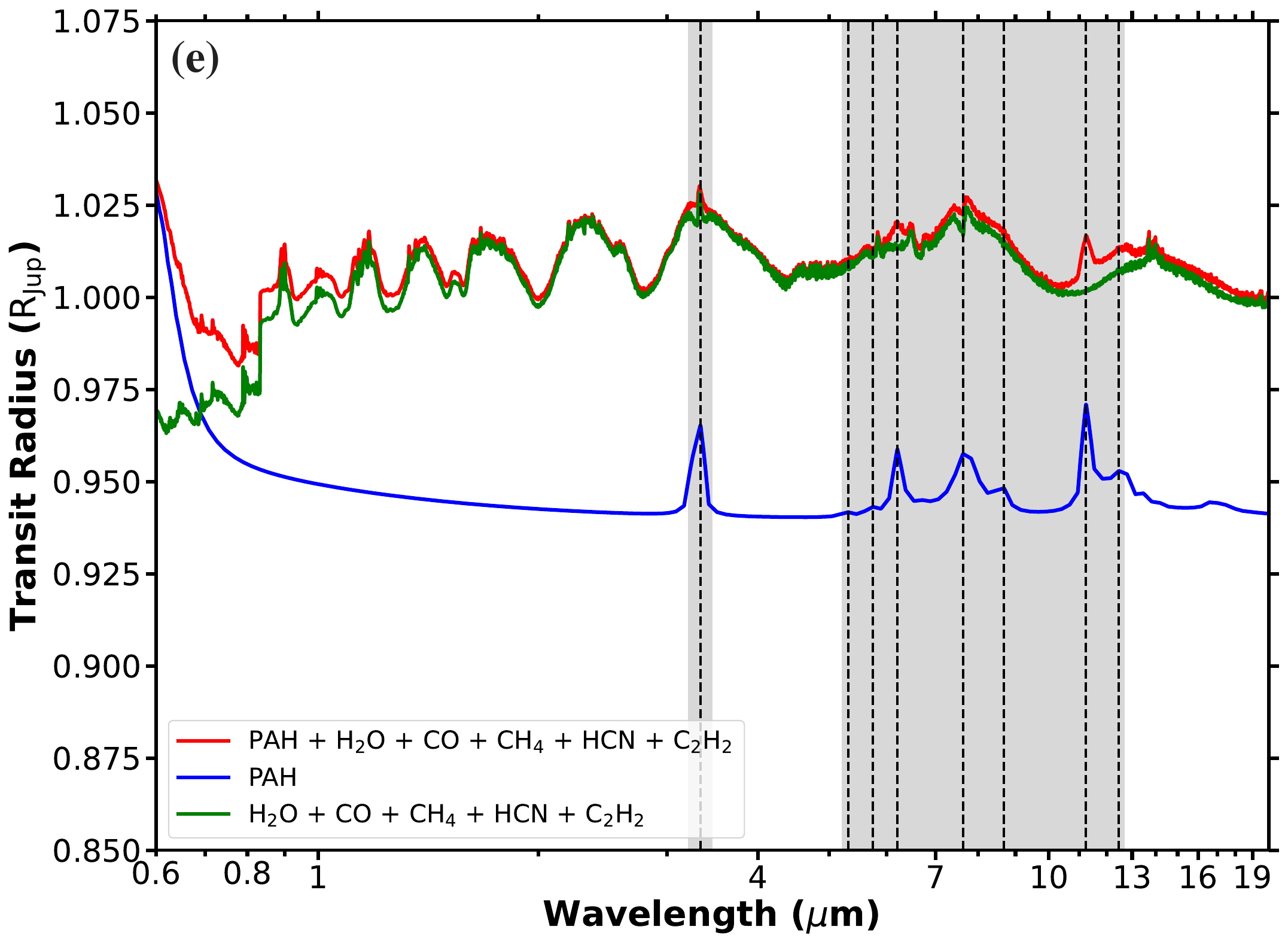}
        \end{minipage}
\caption{Transmission spectra for a modeled planet revolving around a G5 star ($T\mathrm{_{eff}}$ = 1300K, $T\mathrm{_{int}}$ = 200K, [Fe/H] = 1.0, log($g$) = 3.0) with different C/O ratios: (a) C/O = 0.55 (solar) (b) C/O = 0.80 (c) C/O = 1.00 (d) C/O = 1.25 (e) C/O = 1.50. The blue line indicates the PAH cross-section for the abundance obtained from the models for 4 PAHs together. Red and green lines represent transmission spectra with and without considering PAHs, respectively, in the atmosphere. The gray zones cover the areas with the most important PAH features (represented with the dotted lines) at 3.3 $\mu$m and 6-12 $\mu$m.}
\label{fig:transmission_spectra}
\end{figure*}

\begin{figure}
	\includegraphics[width=\columnwidth]{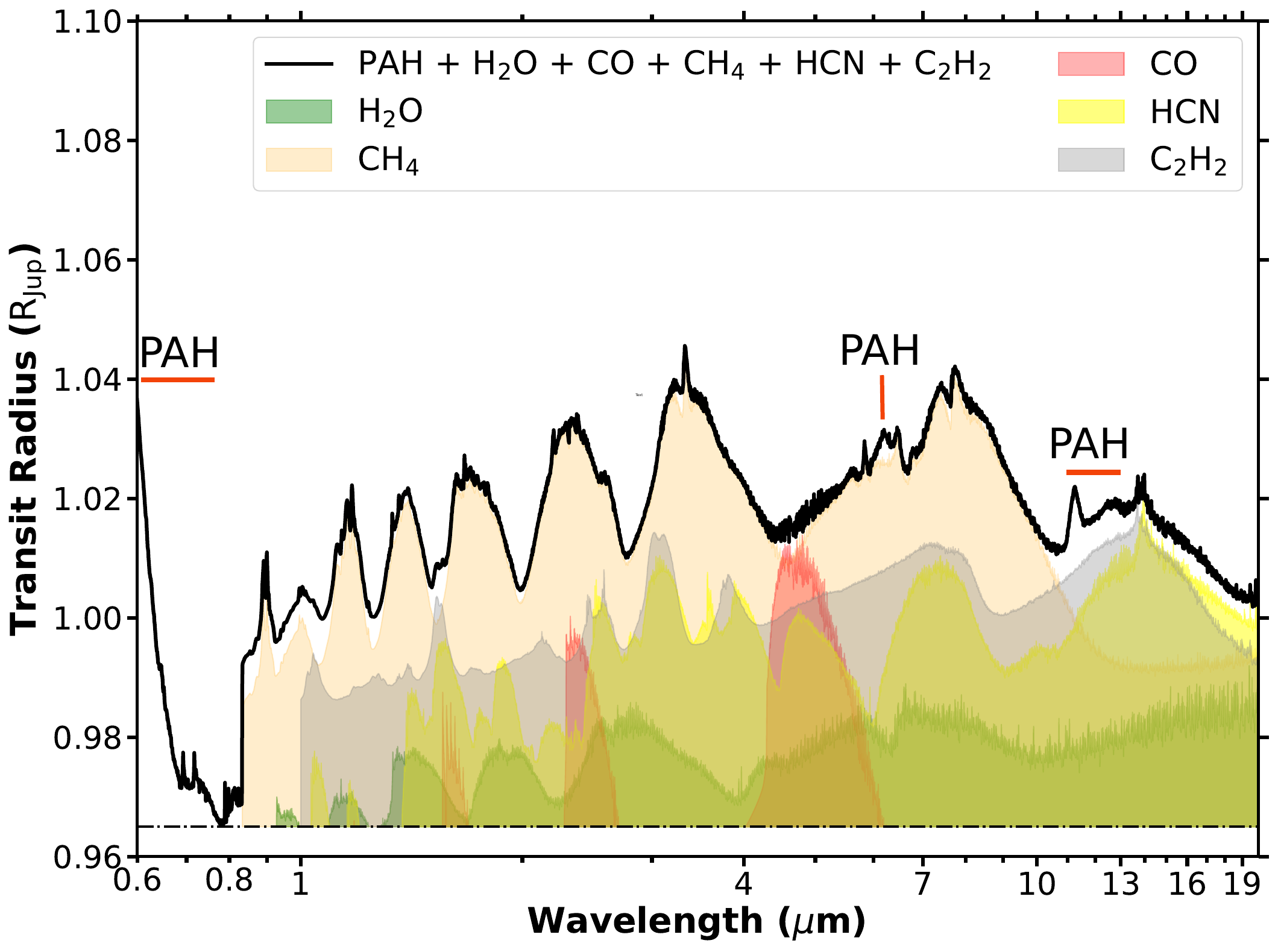}
        \caption{Transmission spectrum for the synthetic planet ($T\mathrm{_{eff}}$ = 1300K, $T\mathrm{_{int}}$ = 200K, C/O = 1.25, [Fe/H] = 1.0, and log($g$) = 3.0) with key contributors to the atmospheric spectrum.}
        \label{fig:contribution}
\end{figure}

The C/O ratio and metallicity have very strong effects on the atmospheric characteristics of a planet. In our models, PAH formation also shows a strong dependency on the C/O ratio. We find that PAH production is low for solar or slightly higher than solar C/O ratios, as the results for C/O $\leq$ 0.80 shown in Figure \ref{fig:VMR_transiting_CtoO}. PAHs in planet atmospheres are more pronounced at lower pressures, primarily attributed to the TP structure. This correlation is shown in Figure \ref{fig:TP_transit}, wherein lower pressures correspond to temperatures that coincide with the optimal range for the formation of PAHs, as illustrated in Figure \ref{fig:0D_model}. The contribution of PAH features to the planet's spectra begins to be visible in models where C/O = 1, which we assume to be the lower limit for thermalized conditions (see Figure \ref{fig:transmission_spectra}). The 3.3 $\mu$m feature is not discernible in the transmission spectra due to the presence of the 3.4 $\mu$m feature of aliphatic hydrocarbon (e.g. $\mathrm{CH_4}$ for our model). It becomes noticeable however as the C/O ratio increases. The pronounced steep slope observed in the UV region (see Figure \ref{fig:transmission_spectra}c, \ref{fig:transmission_spectra}d, and \ref{fig:transmission_spectra}e) results from the adopted PAH cross sections. This observation aligns with the likelihood that photochemical hazes play a significant role in generating super-Rayleigh slopes in the planet's transmission spectra, as highlighted in the study by Ohno et al. (\citeyear{Ohno_2020}). Figure \ref{fig:contribution} illustrates the contributions of various molecules (for a super solar C/O ratio) to the overall planetary spectrum, where $\mathrm{CH_4}$ and $\mathrm{C_2H_2}$ emerge as the primary constituents (the cutoff of $\mathrm{CH_4}$ and $\mathrm{C_2H_2}$ opacity at $\sim$ 0.8 and $\sim$ 1 $\mu$m respectively is not real and is caused by the lack of data). While the absorption features between 6 and 12 $\mu$m, including the spectra slope at UV, appear more prominent, a more detailed investigation of atmospheric retrieval is needed to assess its detectability with current and future space facilities. This will be the focus of Paper II of this series (Grübel et al., in prep.).

The average temperature of the photosphere (between 1-10$^{-3}$ bar pressure range) also allows a better understanding of the 0-D models (see Figure \ref{fig:TP_transit}). An average photospheric temperature $T\mathrm{_{photo}}$ $\sim$ 1300K is obtained from different models. Here, we extended the C/O ratio to 2.0 to better understand the TP profile and to establish a connection between the C/O ratio and PAH formation. The increase of $T\mathrm{_{photo}}$ of the planet with increasing the C/O ratio is not very significant, and hence, the C/O ratio has a dominating role over the planetary effective temperature for the formation of PAHs.

\begin{figure}
	\includegraphics[width=\columnwidth]{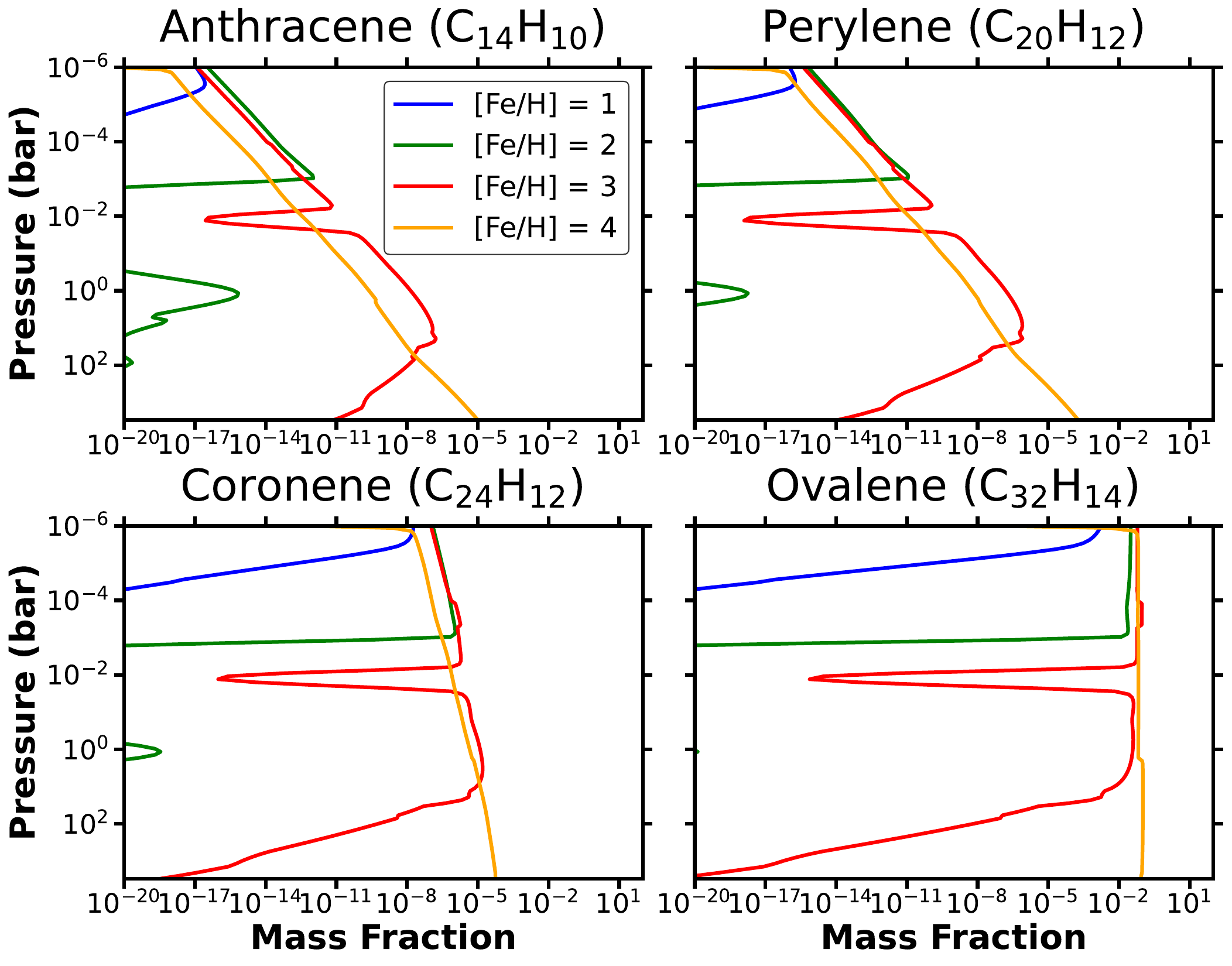}
  \caption{Mass fraction of 4 different PAHs for transiting planet ($T\mathrm{_{eff}}$ = 1300K, $T\mathrm{_{int}}$ = 200K, C/O = 1.0, and log($g$) = 3.0) for different metallicity values. \textbf{(Top Left)}: Anthracene. \textbf{(Top Right)}: Perylene. \textbf{(Bottom Left)}: Coronene, and \textbf{(Bottom Right)}: Ovalene. [Fe/H] = 0 yields PAH abundances less than 10$^\mathrm{-20}$ and hence, not significant for planet atmosphere.}
   \label{fig:VMR_transiting_metallicity}
\end{figure}

 We further investigated the sensitivity of the models to our assumed metallicity values. Figure \ref{fig:VMR_transiting_metallicity} reflects the dependency on metallicity for planets bearing supersolar C/O ratio (C/O = 1.0). Metallicity also impacts the formation of PAHs in a similar way as the C/O ratio does: enhanced metallicity favors PAH formation. However, for Solar C/O ratio, we find that the PAH abundance does not change significantly with the increase of the metallicity, leading us to the conclusion that the C/O ratio is the dominant parameter for the PAH abundance. 

Our current models consider a cloud-free context. However, it is essential to note that the same scenario may not apply to cloudy atmospheres. Specifically, under supersolar C/O ratios or high metal-rich conditions, graphite condensation can exert a substantial influence on atmospheric chemical composition by extracting carbon from the gas phase (see Lodders \& Fegley \citeyear{1997AIPC..402..391L}; Moses et al. \citeyear{moses2013compositional}). As a result, this process may lead to a reduced abundance of PAHs.

At lower C/O ratios, equilibrium models favor the formation of water over any carbon-bearing species (Molaverdikhani et al. \citeyear{Molaverdikhani_2019}). Equilibrium models thus exclude the presence of PAHs at solar C/O. If a transiting planet bears the signature of PAH in its spectra, two possible scenarios can explain that case: (1) various astronomical pathways have led to the incorporation of PAHs into the planetary atmosphere (see Section \ref{sec:ISM}), or (2) the PAHs were formed in-situ, via a chemical pathway that has not been explored in this work. While this study addresses the thermochemical equilibrium scenarios, disequilibrium aspects for in-situ formation will be the focus of future work. 

\begin{figure}
	\includegraphics[width=\columnwidth]{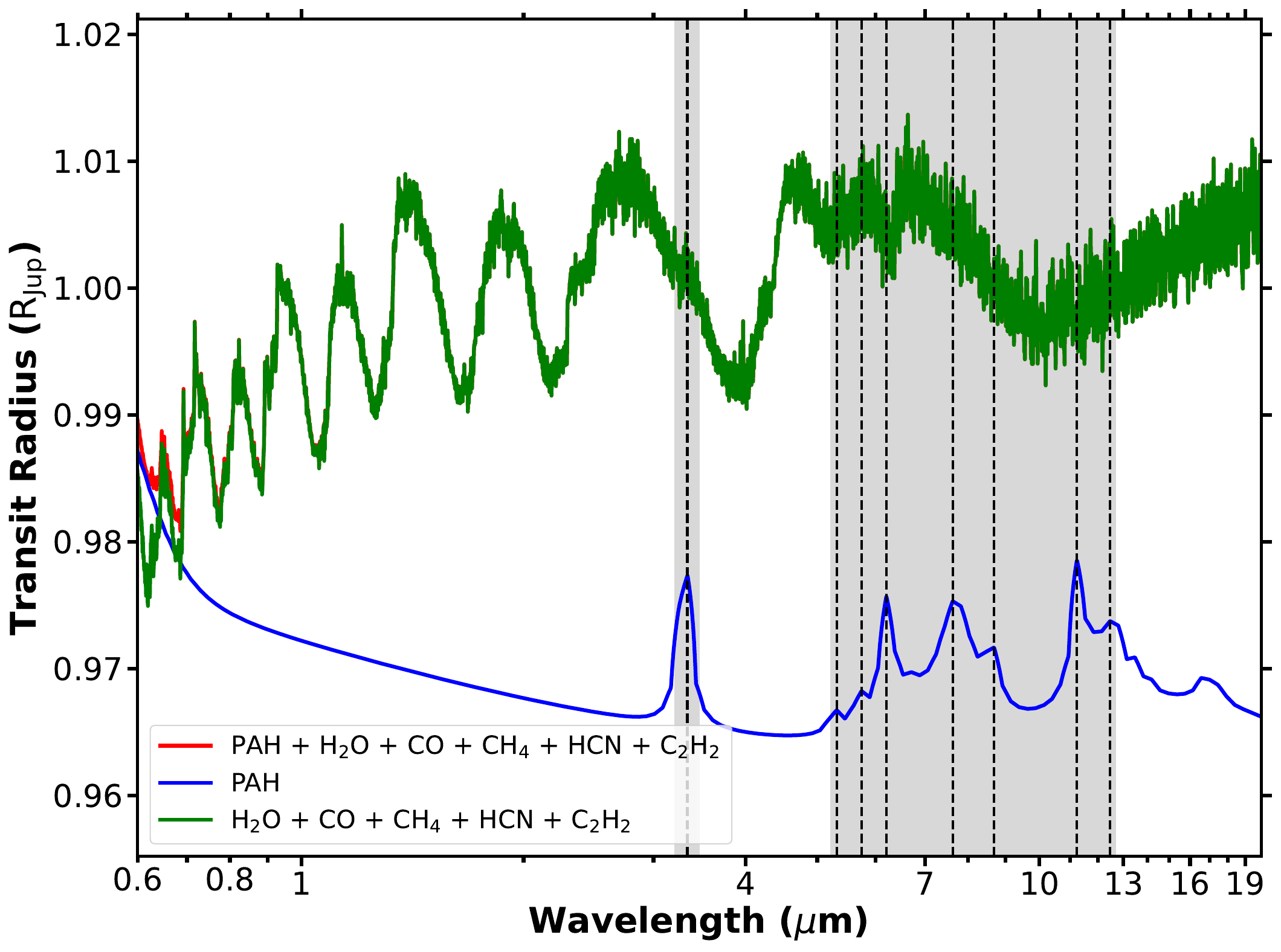}
  \caption{Transmission spectrum for a modeled planet containing ISM abundance for PAHs: $T\mathrm{_{eff}}$ = 1300K, $T\mathrm{_{int}}$ = 200K, [Fe/H] = 1.0, and log($g$) = 3.0.}
   \label{fig:transmission_ISM}
\end{figure}

Figure \ref{fig:transmission_ISM} shows the planetary transmission spectrum for solar C/O ratio where ISM abundance of PAH is considered. Due to the prominence of the $\mathrm{H_2O}$ feature in the spectrum, PAH characteristics are not visible to the naked eye except for the slope. While the PAH features are hidden by the $\mathrm{H_2O}$ features and are not easily discernible by visual inspection, these might be detectable by means of atmospheric retrieval.

%%%%%%%%%%%%%%%%%%%%%%%%%%%%%%%%%%%%%%%%%%%%%%%%%%%%%%%%%%%%%%%%%%%%%
\subsubsection{Directly Imaged Planet}
\label{result:direct_imaged}

To better understand the influence of metallicity and C/O ratio on PAH formation, we extended our investigation to directly imaged planets. Similar to transiting planets, Solar C/O was not able to form PAHs significantly, even with enhanced metallicity values. Therefore, the main investigation was executed for a fixed supersolar C/O ratio but for different metallicities. 

\begin{figure}
	\includegraphics[width=\columnwidth]{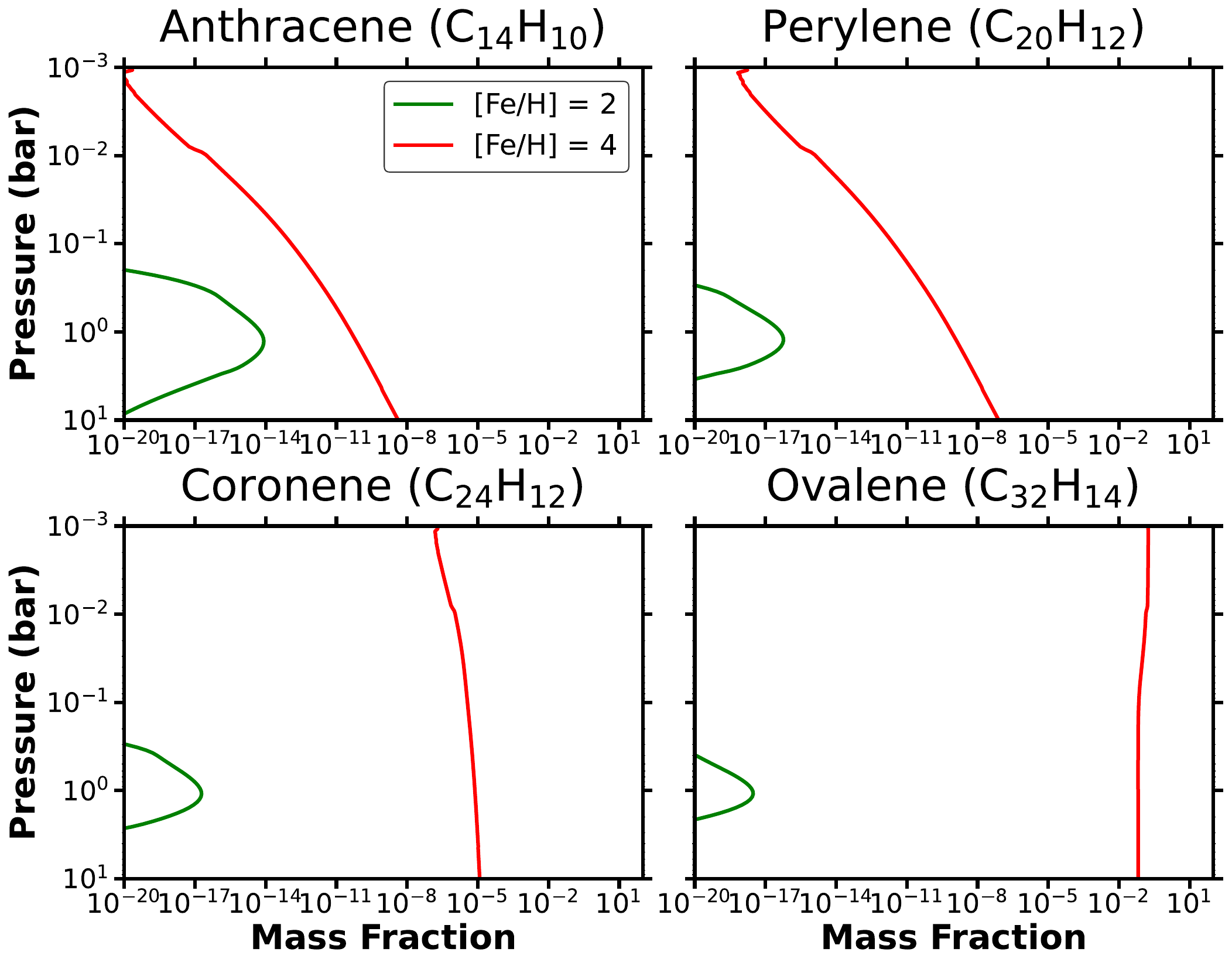}
  \caption{Mass fraction of 4 different PAHs for \textbf{directly} imaged planet model (d = 10au, $T\mathrm{_{int}}$ = 700K, C/O = 1.2, log($g$) = 3.0) for different metallicities. \textbf{(Top Left)}: Anthracene. \textbf{(Top Right)}: Perylene. \textbf{(Bottom Left)}: Coronene, and \textbf{(Bottom Right)}: Ovalene. Similar to Figure \ref{fig:VMR_transiting_metallicity}, we kept a lower limit of 10$^{\mathrm{-20}}$ for molecular abundance. [Fe/H] = 0 and 1 cases produce an insignificant amount of PAHs in the atmosphere and hence, not shown here.}
   \label{fig:VMR_direct_imaged}
\end{figure}

Figure \ref{fig:VMR_direct_imaged} shows the mass fractions of 4 different PAHs at different metallicity values. [Fe/H] = 0 and 1 cases are less important in this regard as they could not produce PAHs significantly. Increasing metallicity boosts PAH formation by a few orders of magnitude, emphasizing the importance of metallicity on PAH formation for directly imaged planets. It is interesting to note that lower metallicities favor the formation of small PAHs, while the formation of larger PAHs is observed for higher metallicity values.

\begin{figure}
	\includegraphics[width=\columnwidth]{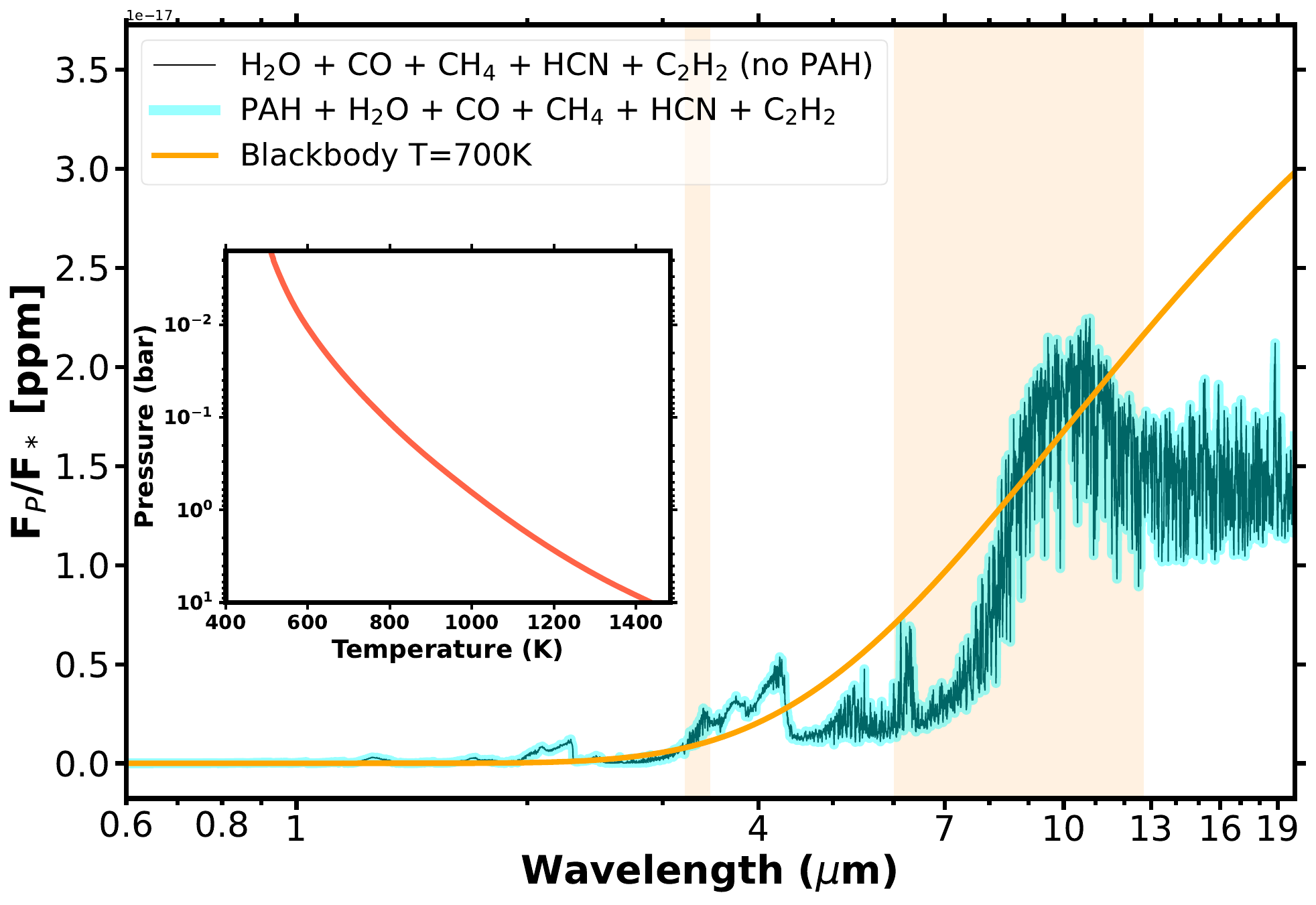}
  \caption{Emission spectrum for a synthetic directly imaged planet: d = 10au, $T\mathrm{_{int}}$ = 700K, C/O = 1.2, log($g$) = 3.0, and [Fe/H] = 2. The orange line represents the Blackbody spectrum for 700K, which is $T\mathrm{_{eff}}$ of the planet.}
   \label{fig:emission_direct_imaged}
\end{figure}

For directly imaged planets, molecular signatures are probed using the absorption features from the planetary emission spectrum. We simulated the planetary emission spectra for [Fe/H] = 2, and no PAH feature was observed from the spectra (see Figure \ref{fig:emission_direct_imaged}). Here, directly imaged planets pose a challenge in identifying the molecular signature of PAHs in emission spectra due to the presence of prominent $\mathrm{CH_4}$ and $\mathrm{C_2H_2}$ absorption features, coupled with low PAH abundance at lower metallicity values. This scenario holds true for the range of expected metallicity values and only starts to break down when we reach unrealistically high metallicities (i.e. [Fe/H] = 4).

%%%%%%%%%%%%%%%%%%%%%%%%%%%%%%%%%%%%%%%%%%%%%%%%%%%%%%%%%%%%%%%%%%%%%%%%%%%%%%%%%%%%%%%%
%%%%%%%%%%%%%%%%%%%%%%%%%%%%%%%%%%%%%%%%%%%%%%%%%%%%%%%%%%%%%%%%%%%%%%%%%%%%%%%%%%%%%%%
%%%%%%%%%%%%%%%%%%%%%%%%%%%%%%%%%%%%%%%%%%%%%%%%%%%%%%%%%%%%%%%%%%%%%%%%%%%%%%%%%%%%%%%%
\section{Conclusion}
\label{sec:conclusion}

We simulated atmospheres self-consistently for transiting and directly imaged planets around a Sun-like star. We have shown the role of planetary effective temperature on PAH formation from the 0-D model. Planets with an effective temperature of around 1300K are the most promising targets for investigating PAHs. Additionally, higher C/O ratios and increased metallicity are crucial in facilitating PAH formation under thermalized conditions. Among the three parameters considered, the C/O ratio exhibits a dominant influence in this context.

The attainment of supersolar C/O ratios in planet atmospheres can be linked to planet formation pathways. In the absence of an efficient release of volatiles from the core or late enrichment through interactions with planetesimals, gas giants that accumulate a substantial portion of their atmospheres after reaching the pebble isolation mass tend to acquire envelopes characterized by substellar C/H and O/H ratios along with notably high C/O ratio compositions (C/O $>$ 1.0) (see Bosman et al. \citeyear{Bosman_2021} for further details). Giant planet formation through core accretion, involving the accumulation of mm or cm-sized pebbles without substantial core-envelope mixing, can also result in supersolar C/O ratios (Madhusudhan et al. \citeyear{10.1093/mnras/stx1139}). Planet formation beyond the $\mathrm{CO_2}$ snowline, accompanied by disc-free migration, is associated with a C/O ratio of approximately 1. Alternatively, for planets forming between the $\mathrm{H_2O}$ and $\mathrm{CO_2}$ snowlines and subsequently undergoing migration through the protoplanetary disc, pebble drift can raise the C/O ratio from approximately 0.7 to about 1.0 (Booth et al. \citeyear{10.1093/mnras/stx1103}). Moreover, the rainout of oxygen-rich refractory species in the atmosphere can further enhance the atmospheric C/O ratio when C/O $<$ 1, potentially leading the atmosphere towards a C/O $\approx$ 1 scenario (Burrows et al. \citeyear{1999ApJ...512..843B}). Recent studies have found many protoplanetary discs that exhibit C/O $\geq$ 1 (Bergin et al. \citeyear{Bergin_2016}; Kama et al. \citeyear{kama2016volatile}; Miotello et al. \citeyear{miotello2019bright}). These discs could potentially serve as the molecular factory for the formation of PAH-rich worlds.

If a planet acquires an ex-situ abundance of PAHs, atmospheric retrieval could lead to the detection of PAH features from planet spectra. To draw robust conclusions regarding detectability, a comprehensive investigation involving a substantial population of planets is necessary. This can be achieved by leveraging data from current and upcoming space missions, such as JWST, Twinkle, and Ariel, which will significantly enhance our understanding of planetary atmospheres.

We particularly focused on irradiated planets as they exhibit a more intricate interplay of equilibrium and disequilibrium chemistry across various regions of their atmospheres. However, this does not imply that non-irradiated planets do not show such processes. Understanding the formation and destruction processes of large PAHs is crucial to connect the presence of PAHs in temperate planets (like our Earth and Titan), which are potential habitable worlds in the universe. Although photochemistry, vertical mixing, and chemical kinetics are essential in this regard, there are caveats for the complex structures of these large molecules. We hope to address these challenges in our next work.

%%%%%%%%%%%%%%%%%%%%%%%%%%%%%%%%%%%%%%%%%%%%%%%%%%%%%%%%%%%%%%%%%%%%%%%%%%%%%%%%%%%%%%%%
%%%%%%%%%%%%%%%%%%%%%%%%%%%%%%%%%%%%%%%%%%%%%%%%%%%%%%%%%%%%%%%%%%%%%%%%%%%%%%%%%%%%%%%
%%%%%%%%%%%%%%%%%%%%%%%%%%%%%%%%%%%%%%%%%%%%%%%%%%%%%%%%%%%%%%%%%%%%%%%%%%%%%%%%%%%%%%%%

%--------------------------------------------------------------------

\begin{acknowledgements}
    This research was supported by the Excellence Cluster ORIGINS, which is funded by the \emph{Deut\-sche For\-schungs\-ge\-mein\-schaft (DFG\/, German Research Foundation)} under Germany’s Excellence Strategy – EXC-2094 – 390783311 (\texttt{http://www.universe-cluster.de/}).\\ \\

      \textit{Softwares}: Our Python scripts are dependent on the following libraries: \texttt{numpy} (Harris et al. \citeyear{harris2020array}), \texttt{matplotlib} (Hunter \citeyear{4160265}), \texttt{jupyter} (Kluyver et al. \citeyear{soton403913}), \texttt{dill} (McKerns et al. \citeyear{mckerns-proc-scipy-2011}), and \texttt{seaborn} (Waskom et al. \citeyear{Waskom2021}).

\end{acknowledgements}

\bibliographystyle{aa}
\bibliography{references} % if your bibtex file is called example.bib


\end{document}